\documentclass[11pt]{article}

\usepackage[a4paper,margin=1in]{geometry}
\PassOptionsToPackage{table}{xcolor}
\usepackage{graphicx}
\graphicspath{{figures/}}
\usepackage{amsmath,amssymb,amsfonts}
\usepackage{array}
\usepackage{tabularx}
\usepackage{multirow}
\usepackage{booktabs}
\usepackage{float}
\usepackage{needspace}
\usepackage{rotating}
\usepackage{algorithm}
\usepackage{algpseudocode}
\usepackage[title]{appendix}
\usepackage{subcaption}
\usepackage{placeins}
\usepackage{enumitem}
\setlist{nosep,leftmargin=1.4em}
\usepackage[numbers,sort&compress]{natbib}
\usepackage{microtype}
\usepackage[colorlinks=true,allcolors=blue,breaklinks=true]{hyperref}

\newcommand{\parencite}{\citep}
\newcommand{\textcite}{\citet}
\AtBeginDocument{\let\cite\citep}

\newcommand{\pihat}{\hat{\pi}}
\newcommand{\code}[1]{\texttt{#1}}
\newcommand{\Xsq}{X^{2}}
\newcommand{\good}[1]{#1}
\newcommand{\bad}[1]{#1}

\newcommand{\bmhead}[1]{\subsubsection*{#1}}
\newcommand{\backmatter}{}

\makeatletter
\newenvironment{breakablealgorithm}
  {\begin{center}%
     \refstepcounter{algorithm}%
     \hrule height.8pt depth0pt \kern3pt%
     \renewcommand{\caption}[2][\relax]{%
       {\raggedright\textbf{Algorithm~\thealgorithm} ##2\par}%
       \kern3pt\hrule\kern5pt%
     }%
  }%
  {\kern3pt\hrule\relax\end{center}}%
\makeatother

\title{\bfseries Benchmarking Goodness-of-Fit and Calibration Algorithms for
Logistic Regression Classifiers:\\ A Large-Scale Simulation Study under Sparse Data}

\author{%
  Ebrahim Khaled Ebrahim\thanks{Department of Applied Statistics, Faculty of Business,
  Alexandria University, Alexandria, Egypt. \texttt{ebrahimkhaled@alexu.edu.eg}. Corresponding author.}
  \and
  Ahmed El-Kotory\thanks{Department of Applied Statistics, Faculty of Business,
  Alexandria University, Alexandria, Egypt.}%
}
\date{\today}

\begin{document}
\maketitle

\begin{abstract}
Binary logistic regression is among the most widely used classification algorithms, yet a classifier is only trustworthy if its predicted probabilities are well calibrated. The classical checks---the Pearson chi-square and deviance statistics---break down precisely in the modern setting where predictors are continuous and the data are sparse (one covariate pattern per observation). Four decades of research have produced dozens of alternative goodness-of-fit and calibration algorithms, yet practitioners still default to the Hosmer--Lemeshow test because it ships with their software. This paper provides a unified taxonomy and a large-scale, reproducible simulation benchmark; more than twenty tests are implemented in the open-source R package \texttt{ebrahim.gof}. We evaluate them across five covariate distributions and four misspecification scenarios, with $10{,}000$ replications each, measuring both Type~I error and power. Several classical tests prove liberal, rejecting correct models far too often, while others have little power. A compact core---McCullagh, Osius--Rojek, le~Cessie--van~Houwelingen, Stute--Zhu, and the GiViTI calibration test---delivers the best balance of correct size and high power, and is consistently more powerful than the ubiquitous Hosmer--Lemeshow test. A low-birth-weight application reinforces the point: a model with omitted interactions slips past nearly every test, exposed only by pairing sensitive tests with a calibration (reliability) curve. We translate these findings into practical, evidence-based guidance for assessing logistic regression fit.
\end{abstract}


\section{Introduction}
\label{sec:intro}

Binary logistic regression remains the canonical algorithm for modeling a
dichotomous outcome from a set of predictors. It is the default classifier in
clinical risk scores, credit scoring, churn prediction, and countless applied
studies, and it is the probabilistic backbone on top of which many modern
machine-learning pipelines are built \parencite{hosmer2013applied,
agresti2013categorical}. Two distinct questions govern whether such a classifier
is any good. The first is \emph{discrimination}---can the model separate the two
classes? The second, often neglected, is \emph{calibration}---do the predicted
probabilities correspond to observed event frequencies, so that among cases assigned a risk of
$0.30$ roughly $30\%$ experience the event? In the language of
classical statistics, calibration is exactly what a \emph{goodness-of-fit} (GOF)
test evaluates \parencite{steyerberg2010assessing, van2019calibration}. A model
can discriminate well yet be badly miscalibrated, producing confident but wrong
probabilities---an increasingly important failure mode as predicted
probabilities are used to drive automated decisions.

A clarification of terminology is warranted here: throughout this paper, ``calibration algorithms'' refers to procedures that \emph{assess} calibration---that is, they test whether a fitted model is calibrated and quantify any departure. This is distinct from the calibration methods of the machine-learning literature \parencite{niculescu2005predicting, guo2017calibration} (for example, Platt scaling or isotonic regression), whose purpose is to \emph{recalibrate} a model and improve its probability estimates. The algorithms benchmarked in this study are diagnostic tests of fit, not remediation techniques.

\subsection{Why the classical tests fail on modern data}
The two textbook GOF statistics, the Pearson chi-square $\Xsq$ and the deviance
$D$, compare observed and fitted counts over the distinct \emph{covariate
patterns} of the data. Their reference chi-square distribution relies on each
pattern being replicated many times. This assumption holds for purely
categorical predictors, but it collapses as soon as a single continuous
covariate (age, weight, income) enters the model: almost every observation then
has its own unique pattern, the number of patterns $J$ grows with the sample
size $n$ (so-called \emph{sparse} or \emph{ungrouped} data, $J \approx n$), and
the asymptotics underpinning $\Xsq \sim \chi^2_{J-p}$ no longer apply
\parencite{kuss2002, farrington1996}. In this regime the classical tests can be
wildly anticonservative, flagging perfectly good models as misfit.

\subsection{The absence of a consensus recommendation}
The failure of $\Xsq$ and $D$ on sparse data launched a long line of
alternatives. \textcite{hosmer1980goodness} proposed grouping observations into
deciles of predicted risk---the Hosmer--Lemeshow (HL) test that became the de
facto standard. Subsequent work added covariate-space partitioning
\parencite{Tsiatis1980, xie2008increasing}, standardized-Pearson $Z$-tests
\parencite{osius1992normal, mccullagh1985, farrington1996}, link-function score
tests \parencite{stukel1988generalized}, smoothing-based tests
\parencite{le1991goodness, le1995goodness}, calibration tests imported from the
machine-learning and clinical-prediction literature
\parencite{spiegelhalter1986calibration, nattino2014new, harrell2015regression},
and resampling/bootstrap procedures \parencite{stute2002model, BAGofT2019}. By
2026 the literature contained more than \emph{fifty} distinct GOF and
calibration algorithms \parencite{liu2024comprehensive, review2024}.

Despite this abundance, applied researchers overwhelmingly continue to report a
single Hosmer--Lemeshow $p$-value---largely because it is the test built into
their statistical package---even though HL is known to depend heavily on the
grouping choice, to have low power against several alternatives, and to behave
erratically in very large samples \parencite{Hosmer1997, Lai2018, MHL2021}.
Earlier comparison studies have helped but stayed narrow in scope:
\textcite{Hosmer1997} weighed a handful of tests within a single
data-generating framework, \textcite{canary2015comparison} contrasted three
grouping-based tests under two binning rules, and the recent, more
comprehensive study of \textcite{liu2024comprehensive} compared several modern
proposals. What is still missing is a single, neutral, head-to-head,
large-scale simulation comparison---across roughly $25$ candidate tests---that
tells a practitioner \emph{which} algorithm to trust.

\subsection{Contributions}
This paper fills that gap. Our contributions are:
\begin{enumerate}
  \item A concise \textbf{taxonomy} (Section~\ref{sec:taxonomy}) that organizes
        more than fifty GOF and calibration algorithms into six coherent
        families, clarifying what each one actually tests; of these, about 25 form the head-to-head comparison; the remainder are close variants of included tests, require grouped or categorical data, or lack a usable open-source implementation.
  \item A \textbf{large-scale, reproducible simulation benchmark}
        (Sections~\ref{sec:design}--\ref{sec:results}) built on the respected
        \textcite{Hosmer1997} framework, spanning five covariate distributions,
        two canonical misspecifications (each at a slight and a pronounced severity), five sample sizes, and $10{,}000$ replications
        per scenario.
  \item A clear identification of the algorithms to \textbf{avoid}---those that
        are \emph{liberal} (inflated Type~I error) or have \emph{near-zero
        power}---and of a robust \textbf{recommended core} that dominates the
        popular Hosmer--Lemeshow test.
  \item A \textbf{real-data demonstration} on the classic low-birth-weight study
        showing why a \emph{panel} of tests and a visual calibration curve should
        accompany any single $p$-value.
  \item An evidence-based \textbf{practical recommendation}
        (Section~\ref{sec:discussion}) for routine model checking.
\end{enumerate}
We deliberately restrict attention to the fully sparse case with continuous
covariates and to standard (unpenalized, low-dimensional) logistic regression,
the setting in which the choice of GOF algorithm matters most and is least
settled.

\section{A Taxonomy of Goodness-of-Fit and Calibration Algorithms}
\label{sec:taxonomy}

\subsection{Setup and notation}
Let $(y_i,\mathbf{x}_i)$, $i=1,\dots,n$, be independent observations with
$y_i\in\{0,1\}$ and let the fitted logistic model give predicted probabilities
\begin{equation}
  \pihat_i \;=\; \frac{\exp(\mathbf{x}_i^{\top}\boldsymbol\beta)}
                     {1+\exp(\mathbf{x}_i^{\top}\boldsymbol\beta)},
  \qquad p=\dim(\boldsymbol\beta).
\end{equation}
Here $\boldsymbol\beta$ is the vector of fitted regression coefficients. The Pearson residual is $r_i=(y_i-\pihat_i)/\sqrt{\pihat_i(1-\pihat_i)}$, and the
Pearson and deviance statistics are
\begin{equation}
  \Xsq=\sum_{i=1}^{n} r_i^{2},
  \qquad
  D = -2\sum_{i=1}^{n}\Big[y_i\log\pihat_i+(1-y_i)\log(1-\pihat_i)\Big].
\end{equation}
Under \emph{grouped} data ($J$ distinct covariate patterns, each replicated $m_i$ times) these
follow $\chi^2_{J-p}$; under \emph{sparse} data ($m_i=1$, $J\approx n$) that
reference is invalid, which is the entire reason the alternatives below exist \parencite{agresti2013categorical, kuss2002}.

\subsection{Six families of algorithms}
We group the algorithms by the mechanism they use to recover a valid reference distribution. This classification is our own, inspired by \textcite{Hosmer1997} and \textcite{GAM2021}, and adopted purely for convenience and to make the large set of tests easier to study in a structured way; it is not a unique or canonical scheme, and the same algorithms could reasonably be organized under other taxonomies. Table~\ref{tab:taxonomy} summarizes the field.

Examining these families reveals a methodological progression: successive tests typically address a specific limitation of a prior method while retaining its core assumptions.

\begin{description}
  \item[(F1) Partition / grouping tests.] These restore a valid $2\times G$ table by sorting observations into $G$ groups, then comparing how many events each group actually contains with how many the model predicts. For group $g$, $o_g$ is the observed number of events, $e_g$ is the expected number (the sum of the fitted probabilities in the group), $n_g$ is the group size, and $\bar\pi_g$ is the group's mean predicted probability; a Pearson-type statistic sums the squared gaps $(o_g-e_g)^2$ across groups. The \textbf{Hosmer--Lemeshow} $\hat C=\sum_g (o_g-e_g)^2/[n_g\bar\pi_g(1-\bar\pi_g)]$ uses $G{=}10$ deciles of predicted risk and is referred to $\chi^2_{G-2}$---an \emph{empirical} simplification validated by simulation, not classical theory \parencite{hosmer1980goodness}. Every later member repairs one HL weakness: \textbf{Pigeon--Heyse} $J^2$ multiplies each group by a factor $\phi_g$ that rescales the variance for the spread of $\hat\pi_i$ within a group, instead of the rough fixed adjustment HL uses, and gives the same answer when predicted risk is roughly constant within every bin \parencite{pigeon1999b}; \textbf{$\mathrm{mHL}_{\text{large}}$} keeps HL's statistic but refers it to a \emph{noncentral} $\chi^2$ so fit is judged by how large the misfit actually is rather than by sample size, which stops HL from rejecting almost every model once the sample is very large \parencite{MHL2021}; \textbf{eHL} drops the fixed bins entirely: it recalibrates the predicted probabilities with isotonic regression and measures, through a likelihood ratio, how far they fall from the observed outcomes, reporting this as an ``e-value''---a score that stays valid even when more evidence is added later---and converts it to a $p$-value by Markov's inequality, giving $p\le 1/e$ \parencite{Henzi2024SafeHL}. The next two change \emph{what} is binned: \textbf{Xie} clusters the \emph{covariate space} ($k$-means) rather than the risk score, targeting interactions and nonlinearity \parencite{xie2008increasing}; \textbf{Pulkstenis--Robinson} groups first by the categorical covariate patterns and then splits each at its median risk, so it applies only when categorical predictors exist \parencite{pulkstenis2002two}. Finally, \textbf{BAGofT} splits the data in two: on the first (``training'') part it searches for the grouping that best exposes a mismatch between observed and predicted events, and on the second (``test'') part it computes a chi-square statistic for that grouping ($\hat T\to\chi^2_K$), so choosing the grouping on separate data keeps the test's size correct---the grouping cannot be tuned to make a good model fail \parencite{BAGofT2019}.
  \item[(F2) Standardized-Pearson ($Z$) tests.] Standardizing means rescaling a statistic to a known reference distribution. These tests keep the raw Pearson chi-square $\Xsq$ (which sums squared standardized residuals over covariate patterns) but address its invalid reference by subtracting its mean and dividing by its standard deviation, giving $Z=(\Xsq-\mathrm{E}[\Xsq])/\sqrt{\widehat{\mathrm{Var}}[\Xsq]}\sim N(0,1)$ under a correct model; recall that $m_i$ is the number of trials at covariate pattern $i$, so $m_i=1$ is the fully sparse case. The family differs only in \emph{how} the variance is obtained. \textbf{Osius--Rojek} estimates it \emph{empirically}, as the residual sum of squares (RSS) of an auxiliary weighted least-squares regression of the residual weights on the design matrix, giving $Z=(\Xsq-(n-p))/\sqrt{A+\mathrm{RSS}}$ with $A=0$ for ungrouped data and a parameter-free mean \parencite{osius1992normal}. \textbf{McCullagh} instead uses the \emph{exact} conditional moments of $\Xsq$ given $\hat{\boldsymbol\beta}$ (analytic higher binomial cumulants weighted by the leverages); it is first-order equivalent to Osius--Rojek but more accurate in the tail \parencite{mccullagh1985}. \textbf{Windmeijer} is the same construction restricted to non-sparse data, trimming extreme $\hat\pi$ \parencite{windmeijer1990}. \textbf{Farrington} adds a first-order correction term to the Pearson statistic itself, chosen to be locally orthogonal to $\hat{\boldsymbol\beta}$ so that the variance is \emph{minimized}---but that very construction makes the variance collapse to zero when every $m_i=1$, so Farrington has \emph{no power under full sparsity} (the reason we exclude it from the power study) \parencite{farrington1996}.
  \item[(F3) Score / link tests.] These embed the fitted model in a slightly richer family (adding extra terms that would be zero if the model were correct) and test those added terms with a score statistic---the slope of the log-likelihood $\mathcal{L}(\boldsymbol\beta)$, evaluated at the fitted coefficients $\hat{\boldsymbol\beta}$, which is near zero when the model fits and far from zero when it does not. The advantage is that the enlarged model never has to be fitted; only its score at the current fit is needed. \textbf{Tsiatis} adds indicator variables for regions of covariate space; its score is a quadratic form in the (observed $-$ expected) regional counts---the conceptual ancestor of every partition test in F1 \parencite{Tsiatis1980}. \textbf{Stukel} adds two shape parameters to a generalized-logistic link and tests link adequacy (asymmetry and tail heaviness) with a $\chi^2_2$ score test \parencite{stukel1988generalized}. The \textbf{information-matrix} test checks the information-matrix equality (Hessian $=-$ outer product of scores), a broad omnibus signal of misspecification \parencite{White1982, Orme1988}.
  \item[(F4) Smoothing tests.] The residual for observation $i$ is $y_i-\hat\pi_i$, the gap between the observed outcome $y_i\in\{0,1\}$ and the fitted probability $\hat\pi_i$; under a correct model these residuals average to zero. These tests smooth the residuals over the covariate range---replacing each by a locally weighted average of its neighbors---and ask whether the smoothed curve departs from zero, which would reveal a systematic local lack of fit that point-by-point statistics miss. The statistic $T$ is standardized so that $\mathrm{E}[T]=1$ under a correct model, because each squared smoothed residual is divided by its own variance. \textbf{le~Cessie--van~Houwelingen (1991)} kernel-smooths the residuals and sums the squared smooths weighted by their inverse variance; smoothing the residuals---which have mean zero---avoids the bias of smoothing the regression function \parencite{le1991goodness}. The \textbf{1995} version re-derives the same idea as an efficient score test for extra variance in a random-intercept model, so it also handles categorical covariates \parencite{le1995goodness}. \textbf{Copas}'s unweighted sum of squares $\sum_i(y_i-\hat\pi_i)^2$ is a special (unsmoothed) case of the 1995 test \parencite{copas1989unweighted}, and variants based on an over-fit \textbf{generalized additive model} (GAM) group on its predicted probabilities for sharper risk separation before applying a classical grouped Pearson statistic \parencite{GAM2021}.
  \item[(F5) Calibration / machine-learning tests.] Imported from clinical prediction and machine learning, these ask directly whether the predicted risk $\hat\pi_i$ matches the observed event frequency---the definition of calibration---working on the fitted probabilities and outcomes rather than on grouped counts. The standard summary fits one extra logistic regression, the calibration model $\mathrm{logit}\,\Pr(Y_i{=}1)=\alpha+\beta\,\mathrm{logit}(\hat\pi_i)$, of the outcome on the model's own linear predictor; a perfectly calibrated model needs no correction, so $\alpha=0$ and $\beta=1$ \parencite{steyerberg2010assessing, van2019calibration}. The two numbers flag different faults: the intercept $\alpha$ (calibration-in-the-large) checks the overall level ($\alpha>0$ predicts too low on average, $\alpha<0$ too high), while the slope $\beta$ checks the spread ($\beta<1$ means the risks are too extreme, the fingerprint of over-fitting, and $\beta>1$ too timid). \textbf{Harrell's unreliability index} is a likelihood-ratio test of the joint hypothesis $(\alpha,\beta)=(0,1)$, so it sees only linear miscalibration and is blind to a curved one \parencite{harrell2015regression}. The \textbf{GiViTI} test removes that blind spot: it replaces the straight line with a flexible polynomial of $\mathrm{logit}(\hat\pi_i)$---the slope-and-intercept model is its degree-one case---likelihood-ratio-tests whether the curve equals the identity, and draws a confidence \emph{belt} showing where on the risk scale the model drifts \parencite{nattino2014new, Finazzi2011}. \textbf{Spiegelhalter}'s $z$-test comes from the Brier score $(1/n)\sum(y_i-\hat\pi_i)^2$: it standardizes the score's calibration component, weighting each residual by $(1-2\hat\pi_i)$ so confident predictions count most, and is $N(0,1)$ under good calibration \parencite{spiegelhalter1986calibration}; needing no bins and reading straight off the probabilities, it is one of the most widely used formal calibration tests in the machine-learning and clinical-prediction literature---for instance, the ``S:z'' check reported by the popular \texttt{rms::val.prob} routine, and a standard check on whether the probability outputs of risk scores and modern classifiers can be trusted. The \textbf{Brier score} itself is a summary measure with no $p$-value of its own \parencite{brier1950}.
  \item[(F6) Bootstrap tests.] When the statistic's null distribution has no usable closed form, these tests approximate it by resampling---repeatedly simulating new outcomes from the fitted model and recomputing the statistic to build an empirical reference. \textbf{Stute--Zhu} forms the cumulative-residual (CUSUM) process marked by the residuals and ordered along the single linear-predictor direction $\hat{\boldsymbol\beta}^\top\mathbf{x}$ (where $\mathbf{x}$ is an observation's covariate vector, $\hat{\boldsymbol\beta}$ the fitted coefficient vector, and $n$ the sample size), and takes a Cram\'er--von~Mises functional, calibrated here by a model-based bootstrap \parencite{stute2002model}. The \textbf{projection} test integrates that directional check over \emph{all} directions for a consistent omnibus test \parencite{liu2024comprehensive} (at a heavy computational cost). \textbf{Lai--Liu} leaves the HL statistic untouched and instead down-samples a huge dataset to a ``standard'' size, so HL's notorious large-$n$ over-power is neutralized \parencite{Lai2018}.
\end{description}

\begin{table}[htbp]
\centering
\caption{A taxonomy of the goodness-of-fit and calibration algorithms compared
in this study. ``Reference'' is the null distribution used to obtain a
$p$-value.}
\label{tab:taxonomy}
\small
\renewcommand{\arraystretch}{1.15}
\begin{tabularx}{\textwidth}{@{}>{\raggedright\arraybackslash}p{2.0cm} >{\raggedright\arraybackslash}X >{\raggedright\arraybackslash}X >{\raggedright\arraybackslash}p{1.9cm}@{}}
\toprule
\textbf{Family} & \textbf{Representative algorithms} & \textbf{Grouping / strategy} & \textbf{Reference} \\
\midrule
F1 Partition      & Hosmer--Lemeshow, Pigeon--Heyse, eHL, $\mathrm{mHL}_{\text{large}}$, Xie, Pulkstenis--Robinson, BAGofT & Bin by predicted risk, covariate space, or an adaptive split & $\chi^2$ \\
F2 Std.\ Pearson  & Osius--Rojek, McCullagh, Windmeijer, Farrington & Standardize $\Xsq$ via corrected moments & Normal \\
F3 Score / link   & Tsiatis, Stukel, Information-matrix (White) & Test extra terms / link curvature & $\chi^2$ \\
F4 Smoothing      & le~Cessie (1991, 1995), Copas, GAM-based & Smooth residuals over covariate range & Normal / $\chi^2$ \\
F5 Calib./ML & Spiegelhalter, GiViTI belt, Unreliability index, Brier & Compare predicted vs.\ observed risk & Normal / $\chi^2$ / score \\
F6 Bootstrap      & Stute--Zhu, Projection, Lai--Liu (Hosmer boot.) & Resample to build the null & Empirical \\
\bottomrule
\end{tabularx}
\end{table}

A central message of the taxonomy is that these algorithms are \emph{not}
interchangeable: a partition test and a smoothing test can disagree on the same
data because they are sensitive to different departures from fit. This is
exactly why an empirical, like-for-like comparison is needed.

\subsection{How the recommended tests work}
\label{sec:algorithms}
So that a reader can \emph{apply} our recommendations without chasing the
original papers, we give a compact algorithm for the Hosmer--Lemeshow baseline (Algorithm~\ref{alg:hosmer_lemeshow}) here; step-by-step procedures for the five recommended core tests
(Section~\ref{sec:results}) are collected in Appendix~\ref{app:algorithms} (Algorithms~\ref{alg:osius_rojek_test}--\ref{alg:giviti}).

\begin{breakablealgorithm}
\caption{\textbf{Hosmer--Lemeshow test ($\hat{C}$)} --- grouped chi-square goodness-of-fit baseline for binary logistic regression \parencite{hosmer1980goodness}. Statistic $\hat{C}\xrightarrow{d}\chi^2_{G-2}$ under a correctly specified model.}
\label{alg:hosmer_lemeshow}
\begin{algorithmic}[1]\small
\Require Data $(y_i,\mathbf{x}_i)_{i=1}^{n}$ with $y_i\in\{0,1\}$; fitted logistic model; number of groups $G$ (default $G=10$); grouping type $\in\{\hat{C},\hat{H}\}$.
\Ensure Statistic $\hat{C}$ and $p$-value.
\State \textbf{Fit / obtain predicted risks:} for each $i$ compute the fitted probability
  $\hat{\pi}_i=\dfrac{1}{1+\exp\!\big(-(\hat\beta_0+\hat{\boldsymbol\beta}'\mathbf{x}_i)\big)}$,
  the maximum likelihood estimate (MLE) plug-in of $\pi(\mathbf{x}_i)=P(Y=1\mid\mathbf{X}=\mathbf{x}_i)$.
\State \textbf{Order the risks:} sort so that $\hat{\pi}_{(1)}\le\hat{\pi}_{(2)}\le\cdots\le\hat{\pi}_{(n)}$.
\If{grouping type $=\hat{C}$ \textbf{(deciles of risk --- data-dependent, equal-sized cutpoints)}}
    \State Partition the ordered $\hat{\pi}_{(i)}$ into $G$ groups of (near) equal size $n_g\approx n/G$;
    \State i.e.\ cutpoints $c_g^{*}=\hat{\pi}_{(\lfloor gn/G\rfloor)}$, and group $g=\{i:\ c_{g-1}^{*}\le\hat{\pi}_i<c_g^{*}\}$.
\ElsIf{grouping type $=\hat{H}$ \textbf{(fixed equal-width cutpoints)}}
    \State Use fixed cutpoints $c_g=g/G$ (e.g.\ $[0,0.1),[0.1,0.2),\dots,[0.9,1]$ for $G=10$);
    \State group $g=\{i:\ c_{g-1}\le\hat{\pi}_i<c_g\}$ (a group may be empty, reducing the effective $G$).
\EndIf
\State $\hat{C}\gets 0$
\For{$g=1$ \textbf{to} $G$}
    \State $n_g\gets \lvert\text{group }g\rvert$ \Comment{number of observations in group $g$}
    \State $o_g\gets \displaystyle\sum_{i\in\text{group }g} y_i$ \Comment{\emph{observed} events (successes) in group $g$}
    \State $\overline{\hat{\pi}}_g\gets \dfrac{1}{n_g}\displaystyle\sum_{i\in\text{group }g}\hat{\pi}_i$ \Comment{mean predicted risk in group $g$}
    \State $e_g\gets \displaystyle\sum_{i\in\text{group }g}\hat{\pi}_i = n_g\,\overline{\hat{\pi}}_g$ \Comment{\emph{expected} events in group $g$}
    \State $\hat{C}\gets \hat{C} + \dfrac{(o_g-e_g)^2}{n_g\,\overline{\hat{\pi}}_g\,(1-\overline{\hat{\pi}}_g)}$
      \Comment{cell contribution; denominator $=\mathrm{Var}$ of $o_g$ under a binomial group}
\EndFor
\State \textbf{Reference distribution:} under $H_0:\ P(Y=1\mid\mathbf{X})=\pi(\mathbf{x})$ (model correct), and with all
  $e_g$ and $n_g-e_g$ sufficiently large, $\hat{C}\ \dot\sim\ \chi^2_{G-2}$.
\State $p\text{-value}\gets 1-F_{\chi^2_{G-2}}(\hat{C})$ \Comment{$F_{\chi^2_{G-2}}$ = CDF of $\chi^2$ with $G-2$ d.f.}
\If{$p\text{-value}<0.05$}
    \State \Return ``Reject $H_0$: evidence of poor fit''
\Else
    \State \Return ``Fail to reject $H_0$: no evidence of poor fit''
\EndIf
\end{algorithmic}
\end{breakablealgorithm}


\section{Simulation Design}
\label{sec:design}

\subsection{Framework}
To make our results directly comparable with four decades of published work, we
adopt the simulation framework of \textcite{Hosmer1997}, which has become the
de facto standard for evaluating logistic-regression GOF tests and underlies
many later studies \parencite{canary2015comparison, liu2024comprehensive,
nygaard2019simulation}. Each test is judged on two criteria:
\begin{itemize}
  \item \textbf{Type~I error (size):} data are generated from a
        \emph{correctly specified} model; a good test rejects at approximately
        the nominal $\alpha=0.05$.
  \item \textbf{Power:} data are generated from a more complex \emph{true} model
        and a \emph{simpler, misspecified} model is fitted; a good test rejects
        often.
\end{itemize}
Across all experiments we use sample sizes $n\in\{200,500,1000,2000,5000\}$,
$10{,}000$ replications per scenario, $\alpha=0.05$, and $G=10$ groups for the
grouping-based tests. The resampling-based Stute--Zhu test uses $B=200$ model-based bootstrap replications per fit, and the full study was parallelized across $24$ logical cores (AMD Ryzen~9 3900X, R~4.4.1). All grouping-based tests are evaluated at the conventional $G=10$; the known sensitivity of the Hosmer--Lemeshow family to this choice (Section~\ref{sec:taxonomy}) is acknowledged in the limitations. With $10{,}000$ replications, the Monte-Carlo standard error of an estimated size near $0.05$ is $\sqrt{0.05\times0.95/10{,}000}\approx0.0022$ (and at most $0.005$ for a power estimate), so differences smaller than about half a percentage point are within simulation noise. Every scenario was seeded reproducibly and run under R~4.4.1 with its default Mersenne--Twister generator. The compared tests use their conventional settings---$G=10$ bins for the grouping-based tests, the default bandwidth holding $\approx\sqrt{n}$ points for le~Cessie, the internal GiViTI belt at forward-selection threshold $q=0.95$, and $100$ resamples for the BAGofT split---as implemented in the \texttt{ebrahim.gof} package and its dependencies (\texttt{givitiR}, \texttt{BAGofT}, \texttt{mgcv}, \texttt{ResourceSelection}), whose versions are pinned in the package \textsc{description} file. Within each replication, every test was applied to the \emph{same} simulated dataset, so all between-test comparisons are paired. As a robustness check we recomputed empirical size at the stricter $\alpha=0.01$ from the stored per-replication $p$-values: the recommended core stayed well-calibrated (mean size $1.0$--$1.8\%$ across the null scenarios---Stute--Zhu and GiViTI essentially exact at $1.0\%$, the standardized-Pearson and smoothing tests only marginally liberal), so the size conclusions do not hinge on the $0.05$ threshold.

\subsection{Type~I error scenarios}
Five covariate settings (Table~\ref{tab:scenarios}) span the practically
important range of predicted-probability shapes, from symmetric
($\text{median}=0.5$) to highly skewed and sparse ($\chi^2(4)$, where rare
events dominate). The skewed, sparse settings ($U(-6,6)$ and $\chi^2(4)$) are the stress tests on which weak algorithms break.

\begin{table}[htbp]
\centering
\caption{Type~I error scenarios (correctly specified model). Quantiles refer to
the induced distribution of $\pihat$. Adapted from \textcite{Hosmer1997}. The multivariate row combines three independent covariates, $x_1\sim U(-6,6)$, $x_2\sim N(0,1.5)$, and $x_3\sim\chi^2(4)$.}
\label{tab:scenarios}
\small
\renewcommand{\arraystretch}{1.2}
\begin{tabular}{@{}l l c c c@{}}
\toprule
\textbf{Covariate distribution} & \textbf{Coefficients} & \textbf{Q1} & \textbf{Median} & \textbf{Q3} \\
\midrule
$U(-6,6)$ & $\beta_0{=}0,\ \beta_1{=}0.8$ & 0.087 & 0.500 & 0.913 \\
$U(-3,3)$ & $\beta_0{=}0,\ \beta_1{=}0.8$ & 0.231 & 0.500 & 0.769 \\
$N(0,1.5)$ & $\beta_0{=}0,\ \beta_1{=}0.8$ & 0.304 & 0.500 & 0.696 \\
$\chi^2(4)$ & $\beta_0{=}-4.9,\ \beta_1{=}0.65$ & 0.025 & 0.062 & 0.202 \\
Multi-Indep.\ ($U,N,\chi^2$) & $\beta_0{=}-1.3,\ \beta_{1,2}{=}0.8/3,\ \beta_3{=}0.65/3$ & 0.204 & 0.386 & 0.608 \\
\bottomrule
\end{tabular}
\end{table}

\subsection{Power scenarios}
We use the two misspecifications most relevant to applied modeling, each at a
\emph{slight} and a \emph{pronounced} level, with $x\sim U(-3,3)$:
\begin{itemize}
  \item \textbf{Omitted quadratic term} (non-linearity): true
        $g(x)=\beta_0+\beta_1 x+\beta_2 x^2$, fitted $g(x)=\beta_0+\beta_1 x$.
        Severity is set through the anchor $\pi(-3){=}Q$ with $Q{=}0.01$
        (slight) and $Q{=}0.4$ (pronounced).
  \item \textbf{Omitted interaction term}: true
        $g(x,d)=\beta_0+\beta_1 x+\beta_2 d+\beta_3 xd$ with $d\sim
        \text{Bernoulli}(0.5)$, fitted without the $xd$ term. Severity is set
        through $I\in\{0.1,0.7\}$.
\end{itemize}
Coefficients are obtained, following \textcite{Hosmer1997}, by solving the
logit-transformed system at fixed anchor probabilities, so that the
\emph{degree} of misfit---not an arbitrary coefficient---is what we control. Concretely, the quadratic scenario fixes $\pi(-1.5)=0.05$, $\pi(3)=0.95$, and $\pi(-3)=Q$, giving $(\beta_0,\beta_1,\beta_2)\approx(-1.138,\,1.257,\,0.035)$ for $Q=0.01$ and $(-3.232,\,0.558,\,0.500)$ for $Q=0.4$; the interaction scenario fixes $\pi(-3,0)=\pi(-3,1)=0.1$, $\pi(3,0)=0.2$, and $\pi(3,1)=0.2+I$, giving $(\beta_0,\beta_1,\beta_2,\beta_3)\approx(-1.792,\,0.135,\,0.270,\,0.090)$ for $I=0.1$ and $(-1.792,\,0.135,\,1.791,\,0.597)$ for $I=0.7$.
(Link-function misspecification is part of the original framework but is beyond
this paper's scope.)

\section{Results}
\label{sec:results}

In classifier terms, the Type~I error rate measures how often a test wrongly condemns a well-calibrated classifier, and power measures how often it catches a miscalibrated one.

\subsection{Type~I error}
\begin{figure}[p]
  \centering
  \begin{subfigure}{\linewidth}
    \centering
    \includegraphics[width=0.84\linewidth]{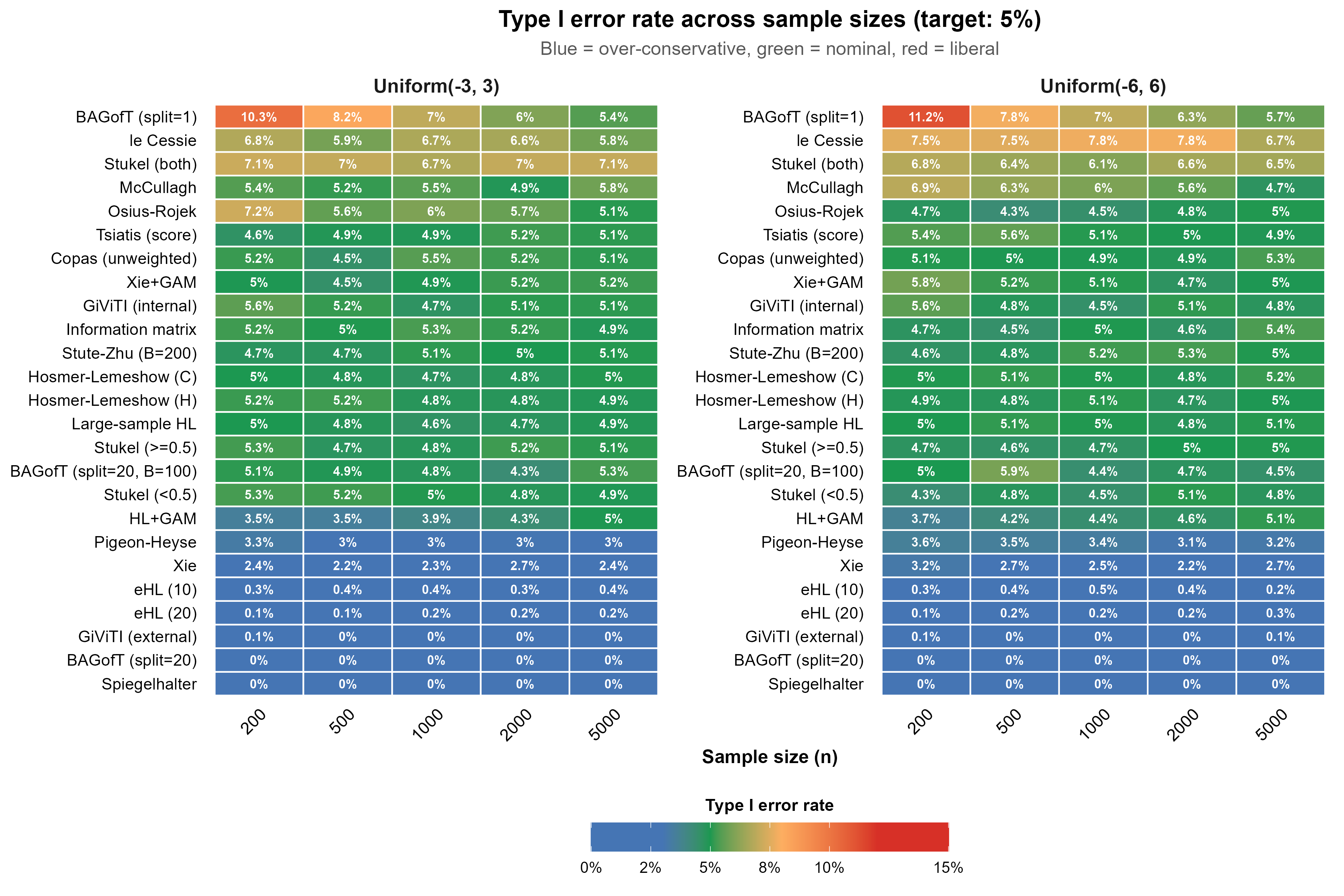}
    \caption{Uniform$(-3,3)$ and Uniform$(-6,6)$ scenarios.}
    \label{fig:type1uniform}
  \end{subfigure}

  \vspace{0.4ex}
  \begin{subfigure}{\linewidth}
    \centering
    \includegraphics[width=0.84\linewidth]{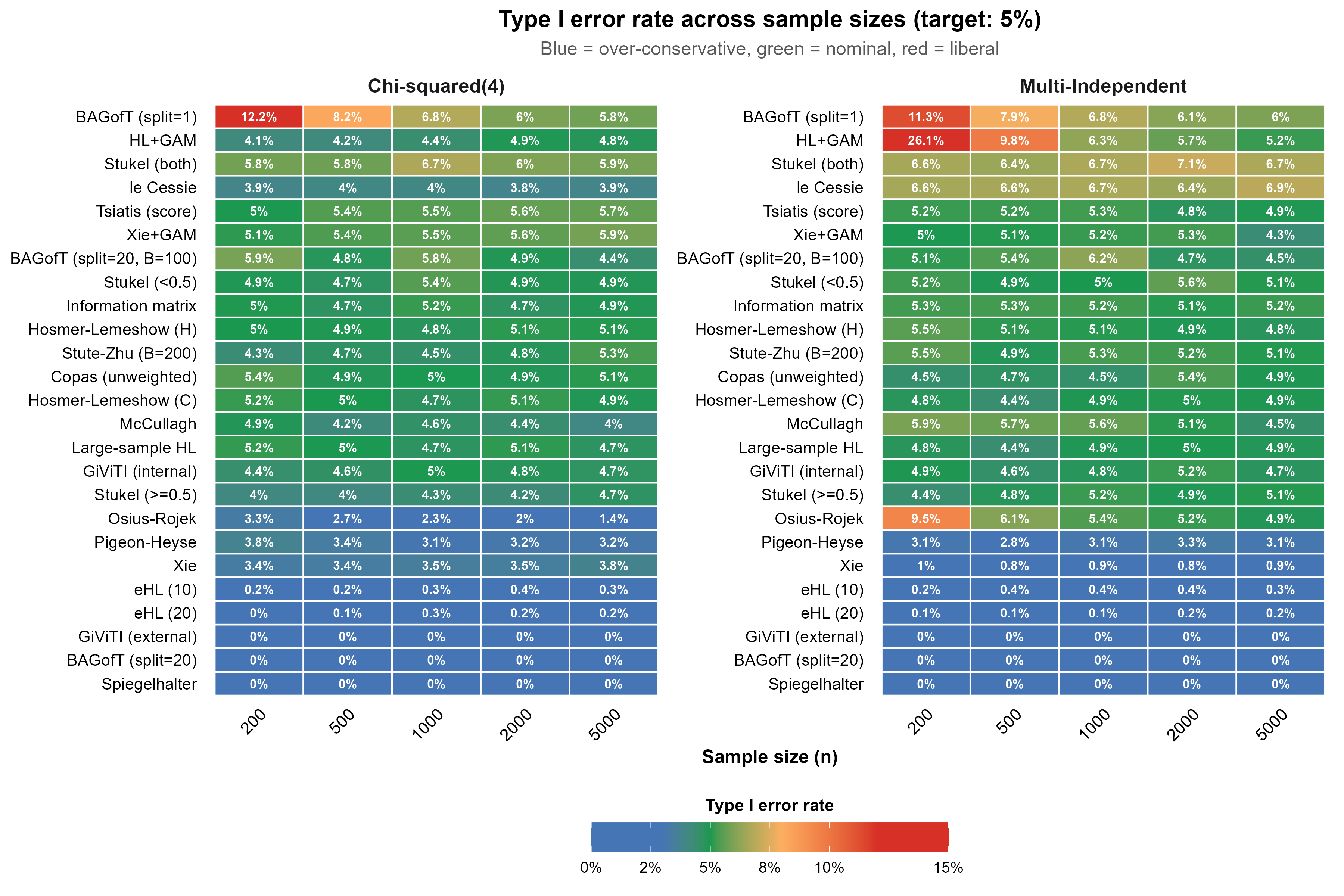}
    \caption{Skewed/sparse $\chi^2(4)$ and Multi-Independent scenarios---the
    hardest stress tests.}
    \label{fig:type1chi2}
  \end{subfigure}
  \caption{Type~I error (empirical rejection rate at $\alpha=0.05$) by test and
  sample size. Values near $0.05$ are well calibrated; larger values indicate
  liberal behavior. In the colour key, blue marks over-conservative tests, green
  the nominal $5\%$ target, and red liberal tests. Under the sparsest scenarios
  (b) several classical tests become strongly liberal, while McCullagh,
  Stute--Zhu, and GiViTI largely retain correct size (Osius--Rojek turns
  conservative under $\chi^2$ skew and liberal at small $n$).}
  \label{fig:type1}
\end{figure}

Figure~\ref{fig:type1} reports empirical
rejection rates under correctly specified models. In the milder settings
($U(-3,3)$, $N(0,1.5)$) most algorithms are well behaved, with the traditional
Hosmer--Lemeshow, Osius--Rojek, the information-matrix test, and McCullagh's
test all hovering near the nominal $5\%$ for $n\ge 500$. The differences emerge
under sparsity ($U(-6,6)$, $\chi^2(4)$, Multi-Independent), and they sort the
field into three groups:

\begin{itemize}
  \item \good{\textbf{Size-robust:}} \code{Stute--Zhu} and the \code{Copas}
        unweighted sum-of-squares test stay close to $5\%$ across \emph{all}
        distributions and sample sizes; \code{Osius--Rojek} is similarly reliable
        but anticonservative at the smallest sample ($n=200$, where its
        $7$--$10\%$ rejection rate in the sparsest scenarios is up to a twofold
        inflation of the nominal $5\%$ level) before converging to the nominal
        level as $n$ grows.
  \item \bad{\textbf{Liberal (inflated size):}} the \code{Pearson},
        \code{Deviance}, Hosmer--Lemeshow \code{$F$-test}, and bootstrap
        \code{Hosmer} tests reject correct models far too often---rejection
        rates climb toward $100\%$ in the sparsest scenarios
        (Section~\ref{sec:excluded}). These tests are unsafe on sparse data.
  \item \textbf{Over-conservative:} \code{Spiegelhalter} and \code{eHL} sit at or
        near a $0\%$ rejection rate. They will rarely raise a false alarm---but,
        as the power analysis confirms, they also miss real problems.
\end{itemize}
As expected, the well-behaved tests converge to the nominal level as $n$ grows.

\subsection{Power: who detects real misfit?}
\begin{figure}[p]
  \centering
  \begin{subfigure}{\linewidth}
    \centering
    \includegraphics[width=0.84\linewidth]{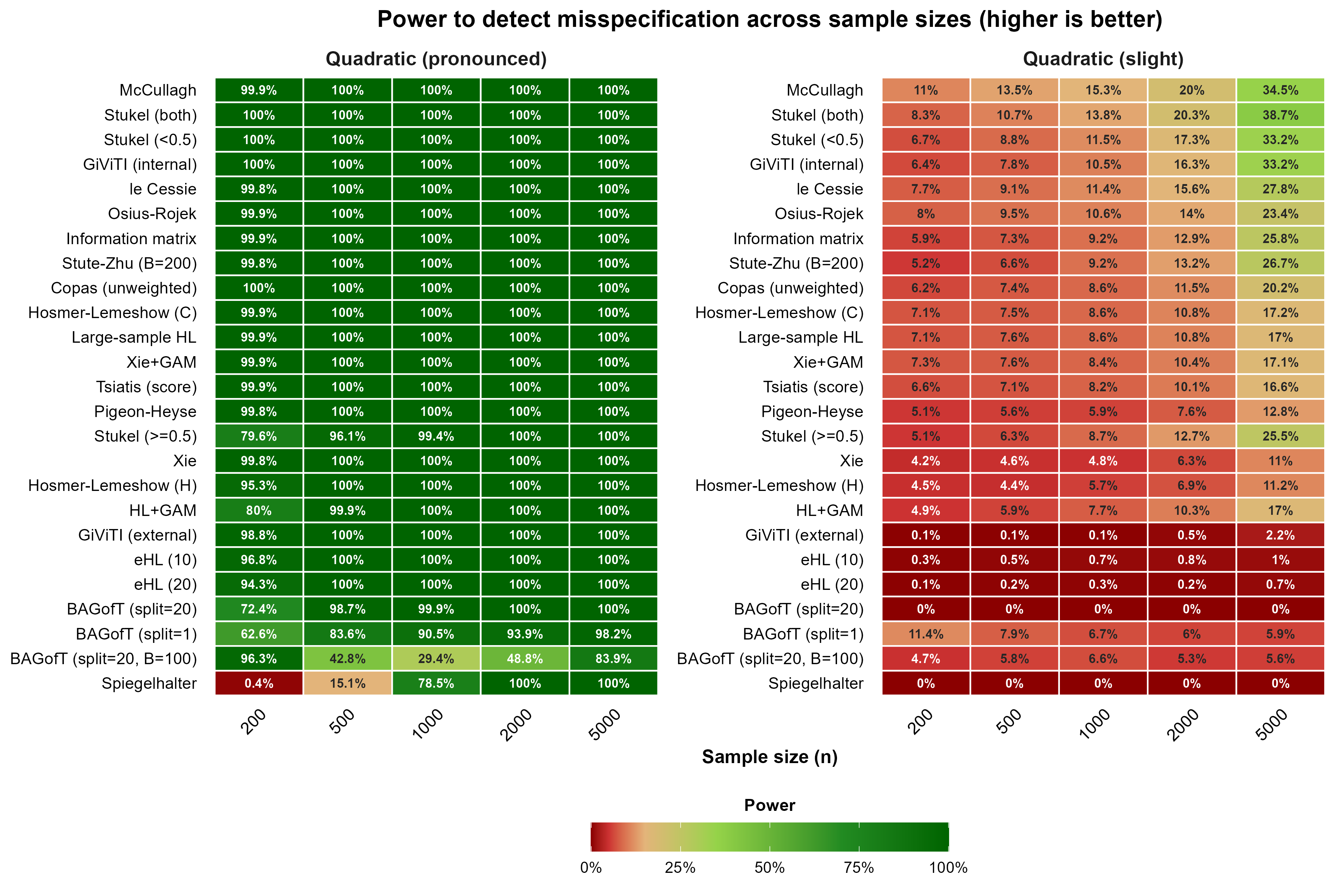}
    \caption{Omitted quadratic term (non-linearity).}
    \label{fig:powerquad}
  \end{subfigure}

  \vspace{0.4ex}
  \begin{subfigure}{\linewidth}
    \centering
    \includegraphics[width=0.84\linewidth]{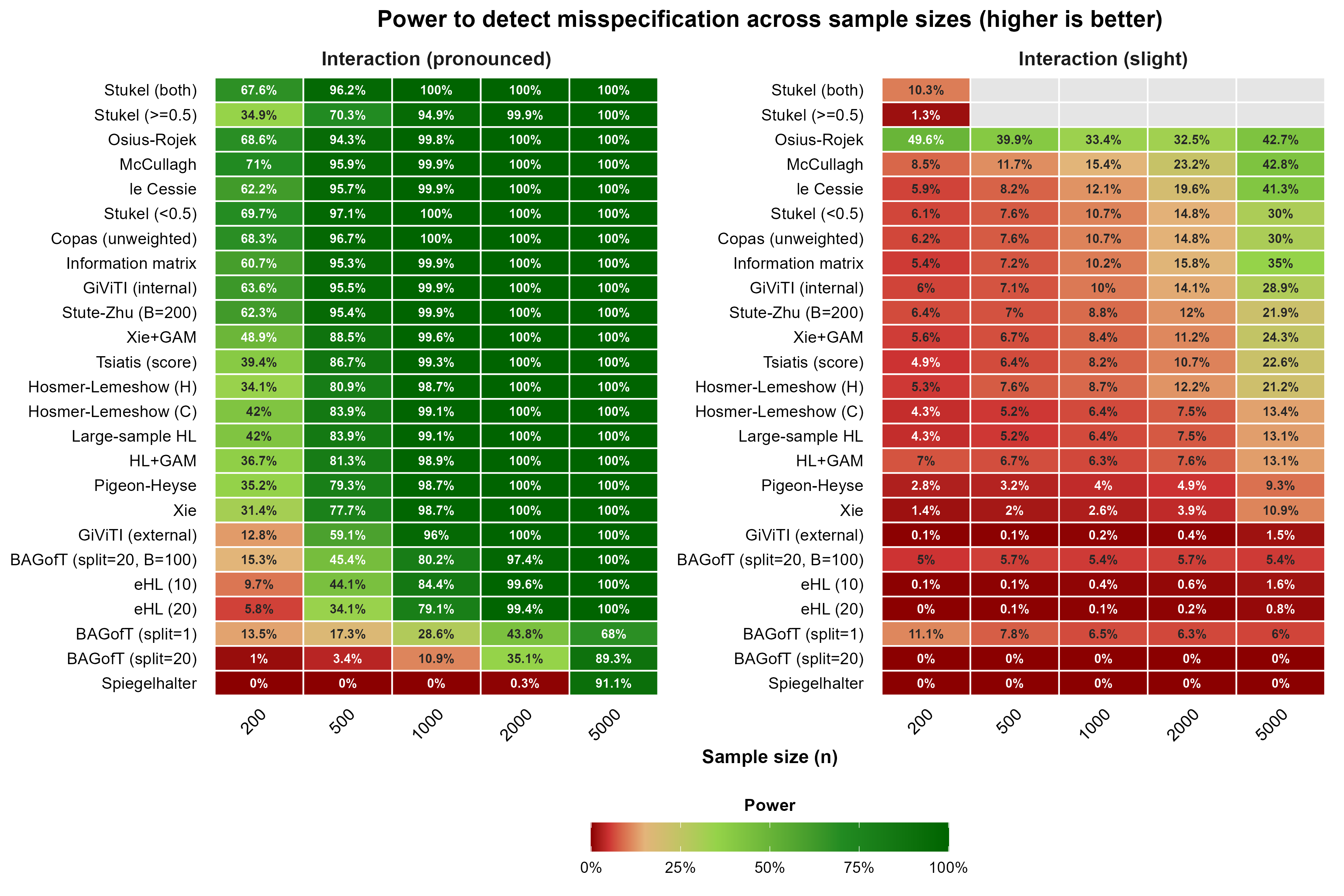}
    \caption{Omitted interaction term.}
    \label{fig:powerint}
  \end{subfigure}
  \caption{Power (empirical rejection rate under misspecified models) by test and
  sample size, for slight and pronounced severities. For the quadratic
  alternative (a) McCullagh, le~Cessie, and GiViTI lead; for the interaction
  alternative (b) McCullagh exceeds $70\%$ power even at $n=200$ under the
  pronounced setting. Grey cells mark tests that returned no value---the Stukel
  refit separates in the slight-interaction scenario.}
  \label{fig:power}
\end{figure}

Figure~\ref{fig:power} reports power against the
omitted quadratic (panel a) and omitted interaction (panel b) misspecifications. When the
misspecification is \emph{pronounced}, most surviving tests reach high power as
$n$ grows; the informative comparison is the \emph{slight} regime, where the
tests separate clearly.

\begin{itemize}
  \item \code{McCullagh}'s test is the standout, achieving over $70\%$ power at
        a sample as small as $n=200$ for the pronounced interaction, and leading
        in the slight regimes.
  \item \code{Osius--Rojek}, \code{le~Cessie}, and \code{Stute--Zhu} are
        consistently among the most powerful, with \code{Stukel} and \code{Copas}
        also strong against the interaction alternative.
  \item The popular \code{Hosmer--Lemeshow} test is mediocre: it controls size
        but trails the standardized-Pearson and smoothing tests in power,
        especially for subtle misfit.
\end{itemize}

\subsection{Algorithms excluded from the main comparison}\label{sec:excluded}
\begin{figure}[tbp]
  \centering
  \includegraphics[width=\linewidth]{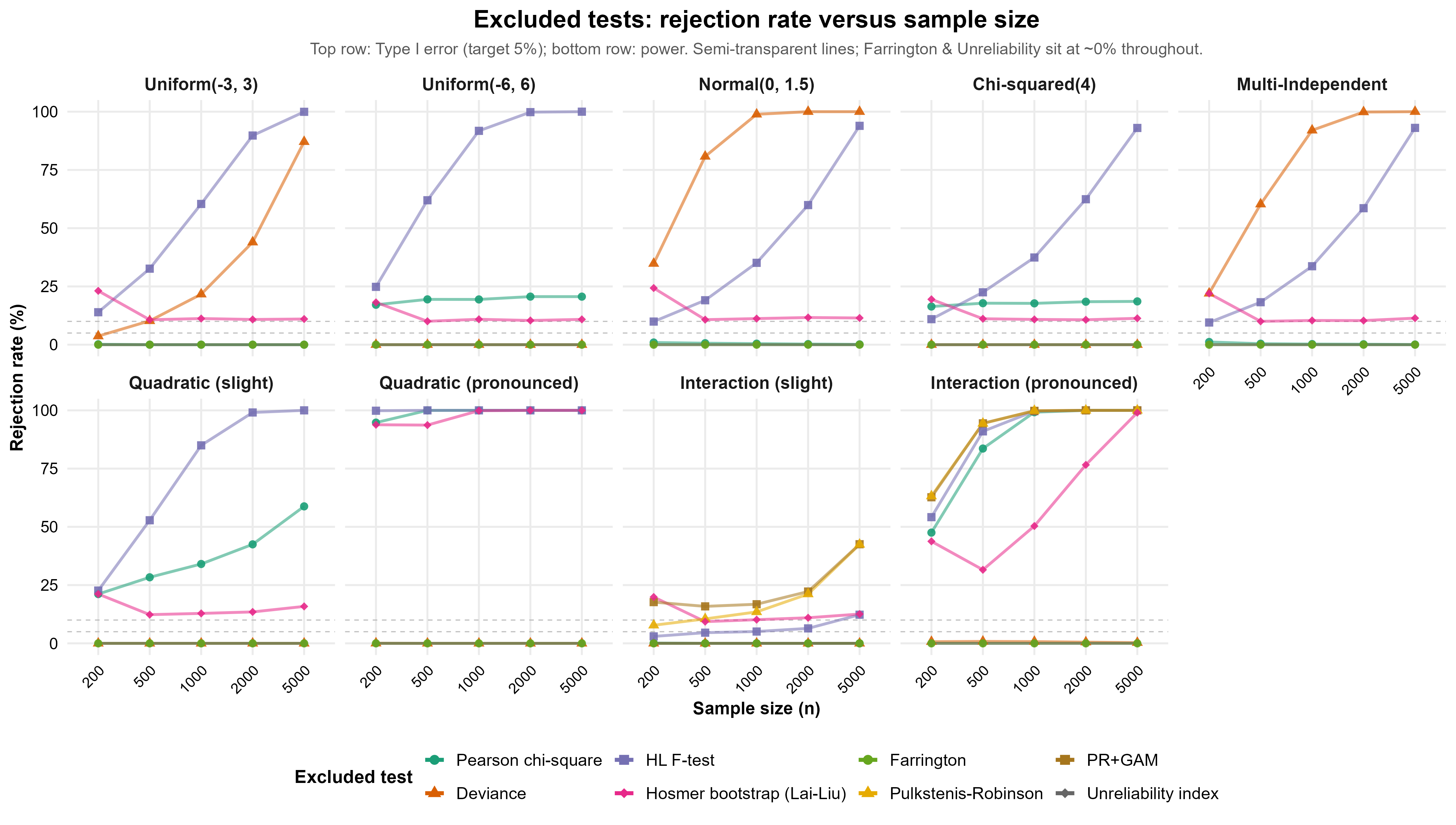}
  \caption{Excluded algorithms. Top row: Type~I error (the classical Pearson,
  deviance, HL $F$-test, and bootstrap Hosmer tests are strongly liberal under
  sparsity). Bottom row: power (the Farrington and unreliability tests have
  essentially zero power).}
  \label{fig:excluded}
\end{figure}

Some algorithms are uninformative in the fully sparse, continuous-covariate
setting and are summarized separately in Figure~\ref{fig:excluded}:
\begin{itemize}
  \item \textbf{Zero power:} the \code{Farrington} test, whose variance collapses
        under full sparsity, essentially never rejects, regardless of how badly
        the model is misspecified. The \code{unreliability index} ($U$) is
        sensitive only to \emph{linear} miscalibration and therefore has no power
        against the (non-linear) alternatives used here.
  \item \textbf{Not generally reliable:} the \code{Pearson} and \code{Deviance} tests, which are sometimes severely liberal and at other times nearly or completely powerless.
  \item \textbf{Not generally applicable:} \code{Pulkstenis--Robinson} requires
        categorical predictors to define its partition, so we exclude it throughout for cross-scenario comparability (only the omitted-interaction scenario carries a binary covariate); the
        \code{projection-based} bootstrap \parencite{liu2024comprehensive} is computationally infeasible at this
        scale.
\end{itemize}

\subsection{Summary of the benchmark}
No single algorithm wins on every axis, but a compact \emph{recommended core}
emerges that combines correct size with high power across the simulated scenarios:
\textbf{McCullagh}, \textbf{Osius--Rojek}, \textbf{le~Cessie--van~Houwelingen},
\textbf{Stute--Zhu}, and the \textbf{GiViTI} calibration test. Membership follows a simple reading of the results: a test enters the core if it maintains its empirical size near the nominal level in \emph{every} null scenario for $n\ge500$ and ranks among the strongest performers in the \emph{slight}-misspecification regimes; GiViTI qualifies as the calibration-family member that satisfies the size criterion while contributing the belt's diagnostic localization of \emph{where} miscalibration occurs. This power advantage is established within our simulation design; as the real-data application in Section~\ref{sec:application} shows, even these tests can fail to detect certain misspecifications (such as an omitted interaction) in genuine covariate structures, which is why we recommend pairing them with a visual calibration curve. Crucially, each of
these is more powerful than the Hosmer--Lemeshow test that dominates applied
practice, while matching or beating its size control.

\section{Real-Data Application: Low Birth Weight}
\label{sec:application}

Simulations show what tests \emph{can} do; a real dataset shows what they
\emph{do} in practice. We revisit the classic low-birth-weight study used by
\textcite{Hosmer1997} and \textcite{kuss2002}, comprising $188$ births, in which $58$ infants had low birth weight ($<2500$\,g).
The outcome is low birth weight and the fitted model uses mother's age
(\code{AGE}), weight at last menstrual period (\code{LWT}), race (two indicators
\code{RACE2}, \code{RACE3}), and smoking status (\code{SMOKE}):
\begin{equation}
  \mathrm{logit}(\pihat) = \beta_0 + \beta_1\,\text{AGE} + \beta_2\,\text{LWT}
  + \beta_3\,\text{RACE2} + \beta_4\,\text{RACE3} + \beta_5\,\text{SMOKE}.
\end{equation}
The fitted coefficients (Table~\ref{tab:lbw}) closely match those of Hosmer and
Lemeshow. Critically, \textcite{Hosmer1997} subsequently identified, through further
modelling efforts on these data, two interactions, $\text{AGE}\times\text{LWT}$ and $\text{SMOKE}\times\text{LWT}$:
the effect of a mother's weight on risk is not constant across age, nor is the
effect of smoking constant across weight. We treat the additive fit as the misspecified working model and these
interaction terms as genuinely omitted structure for illustration---exactly the
kind of subtle misspecification our power analysis targets.

\begin{table}[htbp]
\centering
\caption{Fitted logistic model for the low-birth-weight data
\parencite{Hosmer1997}. The model omits the $\text{AGE}\times\text{LWT}$ and
$\text{SMOKE}\times\text{LWT}$ interactions present in the true model.}
\label{tab:lbw}
\small
\renewcommand{\arraystretch}{1.15}
\begin{tabular}{@{}l r r r r@{}}
\toprule
\textbf{Term} & \textbf{Estimate} & \textbf{Std.\ error} & \textbf{$z$} & \textbf{$p$-value} \\
\midrule
Intercept & $0.2907$  & $1.1127$ & $0.26$  & $0.794$ \\
AGE       & $-0.0204$ & $0.0343$ & $-0.60$ & $0.551$ \\
LWT       & $-0.0127$ & $0.0064$ & $-1.98$ & $0.048$ \\
RACE2     & $1.2735$  & $0.5192$ & $2.45$  & $0.014$ \\
RACE3     & $0.9718$  & $0.4180$ & $2.33$  & $0.020$ \\
SMOKE     & $1.0248$  & $0.3818$ & $2.68$  & $0.007$ \\
\bottomrule
\end{tabular}
\end{table}

\paragraph{Almost every test says ``fine.''}
We applied the full panel of GOF and calibration tests to this misspecified
model. The striking result (Table~\ref{tab:lbwpvals}) is that the overwhelming
majority---including the ubiquitous Hosmer--Lemeshow test ($p=0.26$),
Osius--Rojek ($p=0.36$), McCullagh ($p=0.94$), le~Cessie ($p=0.72$),
Stute--Zhu ($p=0.12$), the GiViTI calibration test ($p=0.59$), and Spiegelhalter
($p=0.87$)---return comfortably non-significant $p$-values and pronounce the
model adequate. Only two \emph{grouped}-data tests detect the omitted
structure: the grouped deviance ($p=0.027$) and the grouped Farrington test
($p=0.013$), with a GAM-based partition borderline (PR$+$GAM, $p=0.069$). (Under full sparsity these grouped statistics are only approximately calibrated, so we read the two detections as sensitive indicators---corroborated by the calibration diagnostics below---rather than as exact tests.) It is no accident that the detectors are grouped tests: in this dataset of $188$ births, $17$ covariate patterns recur two or three times, so although the data are analyzed as fully sparse, they are not perfectly sparse, and a statistic that groups on those repeated patterns is more sensitive to the omitted interaction than the fully sparse form the other tests use. A
practitioner who reported a single Hosmer--Lemeshow $p$-value would have
confidently---and wrongly---concluded that the model fits.

\begin{table}[htbp]
\centering
\caption{Goodness-of-fit and calibration $p$-values on the low-birth-weight
model fitted \emph{without} the $\text{AGE}\times\text{LWT}$ and
$\text{SMOKE}\times\text{LWT}$ interactions. Despite the genuine
misspecification, nearly every test passes the model; only grouped-data tests
flag it (\textbf{bold} = significant at $0.05$).}
\label{tab:lbwpvals}
\small
\renewcommand{\arraystretch}{1.15}
\begin{tabular}{@{}l l c l@{}}
\toprule
\textbf{Test} & \textbf{Family} & \textbf{$p$-value} & \textbf{Verdict} \\
\midrule
Hosmer--Lemeshow            & F1 Partition       & $0.262$ & Adequate \\
Osius--Rojek                & F2 Std.\ Pearson   & $0.361$ & Adequate \\
McCullagh                   & F2 Std.\ Pearson   & $0.943$ & Adequate \\
le~Cessie--van~Houwelingen  & F4 Smoothing       & $0.717$ & Adequate \\
Stute--Zhu (bootstrap)      & F6 Bootstrap       & $0.120$ & Adequate \\
GiViTI calibration          & F5 Calibration     & $0.590$ & Adequate \\
Spiegelhalter               & F5 Calibration     & $0.873$ & Adequate \\
PR $+$ GAM                  & F1/F4 Hybrid       & $0.069$ & Borderline \\
Deviance (grouped)          & Classical       & $\mathbf{0.027}$ & \textbf{Misfit} \\
Farrington (grouped)        & F2 Std.\ Pearson   & $\mathbf{0.013}$ & \textbf{Misfit} \\
\bottomrule
\end{tabular}
\end{table}

\begin{figure}[tbp]
  \centering
  \includegraphics[width=0.78\linewidth]{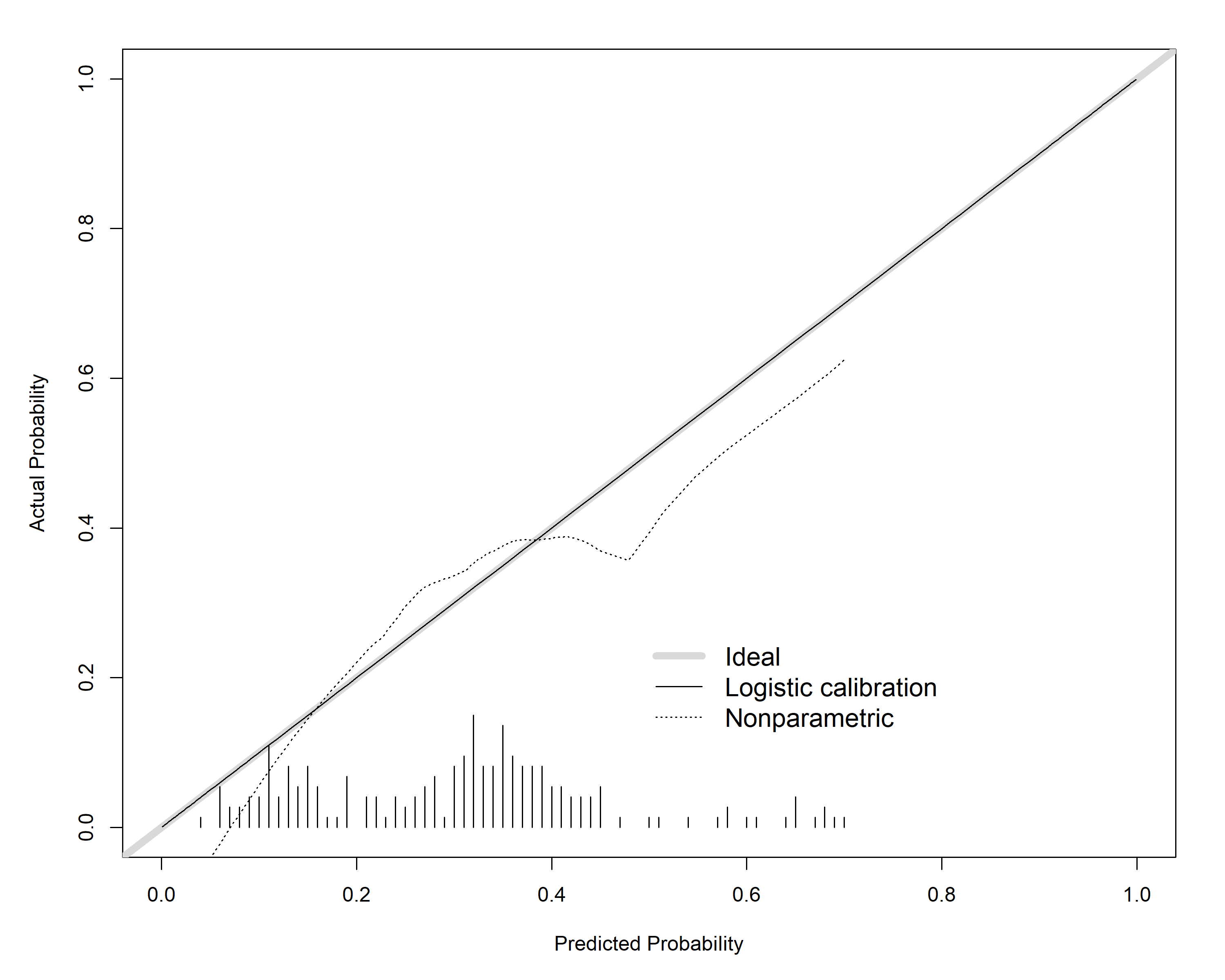}
  \caption{Reliability diagram for the low-birth-weight model (fitted without
  interactions). The solid \emph{logistic-calibration} line sits on the ideal
  diagonal (intercept $\approx 0$, slope $\approx 1$), so linear-calibration
  tests detect nothing; but the dotted \emph{nonparametric} (loess) curve waves
  around the diagonal---dipping below it near predicted risk $0.4$--$0.5$---the
  visible signature of the omitted nonlinearity that most $p$-values miss.
  ($C\approx0.69$, Brier $\approx0.19$.)}
  \label{fig:reliability}
\end{figure}

\paragraph{The visual diagnostic that caught it.}
The early warning came not from a $p$-value but from a \emph{picture}.
Figure~\ref{fig:reliability} is the reliability diagram of the fitted model. Its
\emph{linear} calibration summary looks flawless---intercept $\approx 0$, slope
$\approx 1$, and the solid logistic line lies on the diagonal---which is exactly
why the linear-calibration tests (Spiegelhalter, the unreliability index) see
nothing wrong. But the \emph{nonparametric} (loess) curve tells a different
story: it wanders above and below the diagonal in a wave-like pattern, the
fingerprint of an unmodeled nonlinearity or interaction. Adding the two
interaction terms pulls this loess curve back onto the diagonal and lowers the
calibration error, confirming that the wiggle was real signal, not noise.

\paragraph{The lesson.}
This application crystallizes the paper's central message. First, \textbf{a
single omnibus $p$-value can easily miss real, structured misfit}: here all the
most popular tests---and even the calibration belt---passed a model that is
genuinely incomplete. Second, \textbf{lack of fit is caught by a panel, not a
lone test}: a few sensitive tests together with a visual calibration curve
revealed what the default test could not. Robust assessment of a logistic
classifier means triangulating several diagnostics, never trusting one.

\section{Discussion and Recommendations}
\label{sec:discussion}

Our benchmark yields a simple, actionable picture.

\needspace{4\baselineskip}
\paragraph{What to avoid.} Do not assess the fit of a sparse logistic model
with the raw \textbf{Pearson} or \textbf{deviance} statistic, the
Hosmer--Lemeshow \textbf{$F$-test}, or the \textbf{bootstrap Hosmer} test: all
are liberal and reject correctly specified models far too often. Do not rely on \textbf{Spiegelhalter}
or \textbf{eHL} alone, which are so conservative that they miss real misfit, nor on
the \textbf{Farrington} or \textbf{unreliability} tests, which showed essentially no
power in our scenarios.

\needspace{4\baselineskip}
\paragraph{What to use.} For routine practice we recommend a small \emph{panel}
rather than any single test:
\begin{enumerate}
  \item At least two \textbf{robust omnibus tests} from the recommended core---
        \textbf{McCullagh} and \textbf{Osius--Rojek} (or \textbf{le~Cessie} for
        smooth, local departures, and the score-based \textbf{Stukel} test when link misspecification is the specific concern, though it is not itself part of the recommended core)---which combine correct size with the highest
        power we observed.
  \item A \textbf{visual calibration diagnostic}---a reliability diagram with a
        nonparametric (loess) curve, optionally alongside a \textbf{GiViTI}
        calibration belt---to reveal \emph{where} and \emph{how} the model
        miscalibrates. As our application shows, a wave in the loess curve can
        betray structured misfit that the omnibus $p$-values, and even the belt,
        miss.
  \item \textbf{Context for the $p$-value}: in very large samples even trivial
        misfit becomes ``significant,'' so report effect-size-aware diagnostics
        and treat a lone Hosmer--Lemeshow result with caution
        \parencite{Lai2018, MHL2021}.
\end{enumerate}
The recurring theme is that the algorithm built into the software is not the one
the evidence supports. Hosmer--Lemeshow controls its size but is among the less
powerful options; the standardized-Pearson and smoothing tests it predates are
quietly better. Even the recommended core is not infallible: in small samples, or against complex real-world covariate patterns in sparse data, these tests can still fail to flag genuine misfit---another reason a visual calibration diagnostic is indispensable rather than optional.

\paragraph{Limitations.} Our study deliberately fixes the fully sparse,
continuous-covariate setting, two canonical misspecifications, and $G=10$
groups, and assumes independent, low-dimensional data. Link-function
misspecification, clustered or survey-weighted designs, high-dimensional and
penalized models, and the sensitivity to $G$ are important extensions left to
future work. We also did not separately track quasi-separation of the maximum-likelihood fit in the sparsest small-$n$ cells; the grouped statistics remain computable in such replications, but refit-based procedures may be affected---a caveat shared by all benchmarks in this framework. These restrictions are what make the comparison clean and
reproducible; they also bound its scope.

\section{Conclusion}
\label{sec:conclusion}

Checking that a logistic-regression classifier is well calibrated is as
important as checking that it discriminates, yet the most widely used tool for
the job---the Hosmer--Lemeshow test---is neither the most reliable nor the most
powerful choice on the sparse, continuous-covariate data that dominate modern
practice. Through a unified taxonomy and a large-scale simulation benchmark
built on the \textcite{Hosmer1997} framework, we showed that several classical
tests are dangerously liberal, that several modern ones are powerless, and that
a compact core---McCullagh, Osius--Rojek, le~Cessie--van~Houwelingen,
Stute--Zhu, and the (internal-validation) GiViTI calibration test---offers the best balance of size
and power while outperforming the field's default. A real-data application
reinforced that no single test suffices: a genuinely misspecified model slipped
past nearly every test---including the recommended core---and was caught only by two more-sensitive grouped-data tests and a visual calibration curve. We hope this evidence-based guidance helps practitioners move from
``whatever the software reports'' to a deliberate, trustworthy assessment of model fit.

The practical implication is a change in routine. Assessing the fit of a logistic-regression classifier should not rest on a single test, but on a small panel of robust tests read alongside a visual calibration diagnostic. The $p$-values from these tests answer only whether misfit is \emph{detectable}; a calibration plot shows \emph{where} and \emph{how} the model is wrong. And because no panel is infallible, the calibration plot---not any single $p$-value---remains the final safeguard.

Several questions remain open. Our benchmark fixed the fully sparse, continuous-covariate setting and two canonical misspecifications, so the rankings may shift under link-function misspecification, grouped or moderately sparse data (where Farrington may regain the power it loses under full sparsity), correlated covariates, or high-dimensional and penalized models. Most pressing is the calibration of ``black-box'' learners---gradient boosting, random forests, and neural networks---where trustworthy probabilities matter most yet these tests are largely untested. Extending the recommended core to those settings, and making the most powerful but expensive tests cheaper to compute, are the clearest next steps.

Ultimately, no procedure can guarantee that a given logistic-regression model fits the data well; it can only be interrogated thoroughly and honestly. As predictive models are woven ever more deeply into medicine, finance, and automated decision-making, the cost of an over-confident but miscalibrated classifier rises accordingly. A transparent, multi-faceted assessment---several complementary tests, a calibration curve, and judgment about sample size---is what separates a model that is merely powerful from one that is trustworthy. That is the standard this paper argues should become routine.

\begin{appendices}
\section{Algorithms for the Recommended Core}
\label{app:algorithms}
This appendix collects the step-by-step procedures for the five tests that make up the recommended core (Section~\ref{sec:results}); the Hosmer--Lemeshow baseline is given as Algorithm~\ref{alg:hosmer_lemeshow} in Section~\ref{sec:algorithms}. Algorithms~\ref{alg:osius_rojek_test}--\ref{alg:giviti} are written so that a reader can apply each test without consulting the original papers.

\needspace{4\baselineskip}
\begin{breakablealgorithm}
\caption{\textbf{Osius--Rojek standardized-Pearson test} \parencite{osius1992normal} --- normal-approximation global goodness-of-fit for a logistic regression fit, valid for sparse data / continuous covariates. Statistic: $Z_{O\&R}=\dfrac{X^2-(n-p^{*})}{\sqrt{A+\mathrm{RSS}}}\;\overset{H_0}{\sim}\;N(0,1)$, two-sided.}
\label{alg:osius_rojek_test}
\begin{algorithmic}[1]\small
\Require Fitted logistic model with design matrix $\mathbf{X}\in\mathbb{R}^{n\times p^{*}}$ (including intercept column), binary responses $y_i\in\{0,1\}$, fitted probabilities $\hat{\pi}_i=\big(1+e^{-\mathbf{x}_i^{\top}\hat{\boldsymbol{\beta}}}\big)^{-1}$
\State $n \gets$ number of observations (individual covariate patterns); $\;p^{*} \gets$ number of estimated parameters (rank of $\mathbf{X}$)
\State $v_i \gets \hat{\pi}_i(1-\hat{\pi}_i)$ \Comment{Bernoulli variance of $y_i$; collect $\mathbf{v}=(v_1,\dots,v_n)^{\top}$}
\State $X^2 \gets \displaystyle\sum_{i=1}^{n}\frac{(y_i-\hat{\pi}_i)^2}{v_i}$ \Comment{Pearson chi-square; $E_{H_0}(X^2)=n-p^{*}$}
\Statex \textit{-- Build the Osius--Rojek variance correction via a weighted least-squares (WLS) regression --}
\State $c_i \gets \dfrac{1-2\hat{\pi}_i}{v_i}$ \Comment{Osius weight $w_1(\hat{\pi}_i)=(1-2\hat{\pi}_i)/[\hat{\pi}_i(1-\hat{\pi}_i)]$; pseudo-response, $\mathbf{c}=(c_1,\dots,c_n)^{\top}$}
\State $\mathbf{W} \gets \operatorname{diag}(v_1,\dots,v_n)$ \Comment{WLS weight matrix (weights $=v_i$)}
\State $\hat{\boldsymbol{\gamma}} \gets (\mathbf{X}^{\top}\mathbf{W}\mathbf{X})^{-1}\mathbf{X}^{\top}\mathbf{W}\mathbf{c}$ \Comment{Regress $\mathbf{c}$ on the model covariates $\mathbf{X}$ with weights $v_i$}
\State $\hat{c}_i \gets \mathbf{x}_i^{\top}\hat{\boldsymbol{\gamma}}$ \Comment{WLS fitted values of the pseudo-response}
\State $\mathrm{RSS} \gets \displaystyle\sum_{i=1}^{n} v_i\,(c_i-\hat{c}_i)^2$ \Comment{Weighted residual sum of squares; removes the part of $\mathrm{Var}(X^2)$ explained by estimating $\hat{\boldsymbol{\beta}}$}
\State $A \gets 2\Big(J-\displaystyle\sum_{j=1}^{J}\tfrac{1}{m_j}\Big)$ \Comment{Osius--Rojek's excess-variance term $V_1=2J[1-\mathrm{HM}^{-1}]$, $\mathrm{HM}=$ harmonic mean of the group sizes $m_j$; for pure binary (individual) data every $m_j=1$, so $A=0$ and the denominator reduces to $\sqrt{\mathrm{RSS}}$}
\State $Z_{O\&R} \gets \dfrac{X^2-(n-p^{*})}{\sqrt{A+\mathrm{RSS}}}$ \Comment{Standardized Pearson statistic; denominator $=\hat{\sigma}(X^2)$}
\State $p\text{-value} \gets 2\big(1-\Phi(|Z_{O\&R}|)\big)$ \Comment{$\Phi=$ standard normal CDF (two-sided)}
\State \Return \textbf{reject} $H_0$ (poor fit) if $|Z_{O\&R}|>z_{1-\alpha/2}$, else \textbf{fail to reject}
\end{algorithmic}
\end{breakablealgorithm}

\needspace{4\baselineskip}
\begin{breakablealgorithm}
\caption{\textbf{McCullagh's test} \parencite{mccullagh1985} --- goodness-of-fit for logistic regression via the Pearson statistic standardized by its \emph{conditional} asymptotic moments given the fitted parameters. Statistic: $Z_{\mathrm{McC}}=\dfrac{X^{2}-E(X^{2}\mid\hat{\boldsymbol\beta})}{\sqrt{\mathrm{Var}(X^{2}\mid\hat{\boldsymbol\beta})}}\ \dot\sim\ N(0,1)$ under $H_0$ (good fit).}
\label{alg:mccullagh}
\begin{algorithmic}[1]\small
\State \textbf{Input:} for each covariate pattern $i=1,\dots,n$: observed events $y_i$, number of trials $m_i$ (ungrouped/pure binary: $m_i=1$), fitted probabilities $\hat\pi_i=\mathrm{expit}(\mathbf{x}_i^{\!\top}\hat{\boldsymbol\beta})$ from the logistic fit; design matrix $\mathbf{X}\in\mathbb{R}^{n\times p^{*}}$ with $p^{*}$ the number of estimated parameters.
\Statex \textit{(Conditioning statistic is the sufficient statistic $S^{\alpha}=\mathbf{X}^{\!\top}\mathbf{y}$ for $\boldsymbol\beta$; moments below are conditional on it, valid for large residual d.f.\ even when the $m_i$ are small.)}

\State \textbf{Step 1 --- Pearson statistic.} Compute the standard (uncorrected) Pearson $X^2$:
\Statex \hspace{1em}$X^{2}\gets\displaystyle\sum_{i=1}^{n}\frac{(y_i-m_i\hat\pi_i)^{2}}{m_i\hat\pi_i(1-\hat\pi_i)}$.

\State \textbf{Step 2 --- Binomial cumulants of $y_i$ and weights.} For each $i$ form the 2nd, 3rd and 4th central cumulants of the $\mathrm{Bin}(m_i,\hat\pi_i)$ variate and the working weights:
\Statex \hspace{1em}$\hat\kappa_{i2}\gets m_i\hat\pi_i(1-\hat\pi_i)$ \Comment{variance $=$ weight}
\Statex \hspace{1em}$\hat\kappa_{i3}\gets m_i\hat\pi_i(1-\hat\pi_i)(1-2\hat\pi_i)=\hat\kappa_{i2}(1-2\hat\pi_i)$ \Comment{third cumulant}
\Statex \hspace{1em}$\hat\kappa_{i4}\gets m_i\hat\pi_i(1-\hat\pi_i)\bigl[1-6\hat\pi_i(1-\hat\pi_i)\bigr]$ \Comment{fourth cumulant}
\Statex \hspace{1em}$\mathbf{W}\gets\mathrm{diag}(\hat\kappa_{i2})=\mathrm{diag}\!\bigl(m_i\hat\pi_i(1-\hat\pi_i)\bigr)=\hat{\mathbf V}$.

\State \textbf{Step 3 --- Information matrix, hat matrix, leverages.} Using $\mathbf{W}$ as the iteratively reweighted least squares (IRLS) weight:
\Statex \hspace{1em}$\mathbf{I}^{-1}\gets(\mathbf{X}^{\!\top}\mathbf{W}\mathbf{X})^{-1}$ \Comment{$\approx\widehat{\mathrm{cov}}(\hat{\boldsymbol\beta})$; $\lambda_{ii}\!=\!(\mathbf X\mathbf I^{-1}\mathbf X^{\!\top})_{ii}$}
\Statex \hspace{1em}$h_i\gets(\mathbf{X}\,\mathbf{I}^{-1}\mathbf{X}^{\!\top})_{ii}$ \Comment{$i$th diagonal of the hat matrix $\hat{\mathbf P}=\mathbf X\mathbf I^{-1}\mathbf X^{\!\top}\mathbf W$ is $W_{ii}h_i$}

\State \textbf{Step 4 --- Supplementary (skewness) regression.} Define the standardized-skewness score vector $\mathbf{u}$ with components
\Statex \hspace{1em}$u_i\gets\dfrac{\hat\kappa_{i3}}{\hat\kappa_{i2}^{2}}=\dfrac{1-2\hat\pi_i}{m_i\hat\pi_i(1-\hat\pi_i)}$,
\Statex and regress $\mathbf u$ on $\mathbf X$ with weights $\mathbf W$; the fitted values are
\Statex \hspace{1em}$\hat{\mathbf u}\gets\mathbf{X}\,\mathbf{I}^{-1}\mathbf{X}^{\!\top}\mathbf{W}\,\mathbf{u}$ \Comment{$\hat u_i=\hat\gamma_i$, the ``fitted'' skewness}
\Statex The weighted residual sum of squares of this regression (used for the variance) is
\Statex \hspace{1em}$RSS_u\gets\mathbf{u}^{\!\top}\bigl(\mathbf{W}-\mathbf{W}\mathbf{X}\,\mathbf{I}^{-1}\mathbf{X}^{\!\top}\mathbf{W}\bigr)\mathbf{u}
      =\hat C_3^{\!\top}\hat{\mathbf V}^{-1}(\mathbf I-\hat{\mathbf P})\hat C_3\ \ge 0$,
\Statex where $\hat C_{3,i}=\hat\kappa_{i3}/\hat\kappa_{i2}$ and $\hat{\mathbf P}=\hat{\mathbf V}\mathbf X(\mathbf X^{\!\top}\hat{\mathbf V}\mathbf X)^{-1}\mathbf X^{\!\top}$ (McCullagh eq.~9).

\State \textbf{Step 5 --- Conditional mean $E(X^{2}\mid\hat{\boldsymbol\beta})$.} Correct the naive value $n-p^{*}$ by an $O(1)$ skewness term (McCullagh eq.~6):
\Statex \hspace{1em}$E(X^{2}\mid\hat{\boldsymbol\beta})\gets(n-p^{*})-\dfrac12\displaystyle\sum_{i=1}^{n}\frac{\hat\kappa_{i4}\,h_i}{\hat\kappa_{i2}}+\dfrac12\sum_{i=1}^{n}\hat u_i\,\hat\kappa_{i3}\,h_i$.
\Statex \Comment{The last two terms are the fitted-parameter corrections; for ungrouped binary data they are typically small but non-zero.}

\State \textbf{Step 6 --- Conditional variance $\mathrm{Var}(X^{2}\mid\hat{\boldsymbol\beta})$.} For independent binomial data McCullagh eq.~(9):
\Statex \hspace{1em}$\mathrm{Var}(X^{2}\mid\hat{\boldsymbol\beta})\gets\Bigl(1-\dfrac{p^{*}}{n}\Bigr)\Bigl\{\,2\displaystyle\sum_{i=1}^{n}\frac{m_i-1}{m_i}+RSS_u\,\Bigr\}$.
\Statex \Comment{First term $=0$ for pure binary responses ($m_i\equiv1$); then the conditional variance is driven entirely by $RSS_u$, which can range over several orders of magnitude with the observed configuration --- the key point of McCullagh (1985).}

\State \textbf{Step 7 --- Standardize and test.} Form the normalized statistic
\Statex \hspace{1em}$Z_{\mathrm{McC}}\gets\dfrac{X^{2}-E(X^{2}\mid\hat{\boldsymbol\beta})}{\sqrt{\mathrm{Var}(X^{2}\mid\hat{\boldsymbol\beta})}}$,
\Statex compare to $N(0,1)$: two-sided $p=2\bigl(1-\Phi(|Z_{\mathrm{McC}}|)\bigr)$ (or one-sided upper $p=1-\Phi(Z_{\mathrm{McC}})$).
\Statex \textbf{Decision:} reject $H_0$ (adequate fit) at level $\alpha$ if $|Z_{\mathrm{McC}}|>z_{1-\alpha/2}$.
\end{algorithmic}
\end{breakablealgorithm}

\needspace{4\baselineskip}
\begin{breakablealgorithm}
\caption{\textbf{le~Cessie--van~Houwelingen kernel-smoothed-residuals test} \parencite{le1991goodness} --- an omnibus goodness-of-fit test for logistic regression with continuous covariates, built from a weighted sum of squared Nadaraya--Watson--smoothed Pearson residuals. Statistic $\hat T = n^{-1}\sum_i \hat r_s(X_i)^2\,v(X_i)$; referred, after the estimation correction, to a scaled chi-square $b\,\hat T \sim \chi^2_\nu$ (equivalently a standard normal via $z=(\hat T-\mathrm{E}(\hat T))/\sqrt{\widehat{\mathrm{Var}}(\hat T)}$).}
\label{alg:le_cessie_van_houwelingen_1991_smoothed_residual_test}
\begin{algorithmic}[1]\small
\Require Data $\{(X_i,Y_i)\}_{i=1}^n$, $Y_i\in\{0,1\}$; kernel $K$ (uniform on $[-\tfrac12,\tfrac12]$ by default), standardized bandwidth $h_n$
\Statex \textit{Default $h_n$: chosen so each smoothing region holds $\approx\sqrt{n}$ points; for $d$ covariates use the multiplicative kernel $K(z)=\prod_{l=1}^d K(z_l)$ with per-covariate bandwidth $h_l=h_n s_l$, $s_l=\mathrm{sd}$ of covariate $l$.}
\State Fit the logistic model by ML: obtain $\hat\beta$ and fitted probabilities $\hat g(X_i)=\hat\pi_i=\dfrac{\exp(X_i^{*}\hat\beta)}{1+\exp(X_i^{*}\hat\beta)}$, where $X^{*}=(\mathbf 1,X)$ concatenates the intercept column. \Comment{$H_0:\ g(x)=g_0(x)=g(x,\beta_0)$}
\State Form raw residuals $\hat e_i = Y_i-\hat\pi_i$ and the diagonal variance matrix $\hat V=\mathrm{diag}(\hat v_i)$, $\hat v_i=\hat\pi_i(1-\hat\pi_i)$.
\State Form standardized (Pearson) residuals $\hat r_i = \dfrac{Y_i-\hat\pi_i}{\sqrt{\hat\pi_i(1-\hat\pi_i)}}$, i.e. $\hat{\mathbf r}=\hat V^{-1/2}\hat{\mathbf e}$ (variance $1$ under $H_0$).
\For{$i=1$ to $n$, $j=1$ to $n$}
    \State $w_{ij}\gets K\!\big[(X_i-X_j)/h_n\big]$ \Comment{unnormalized kernel weight of point $j$ when smoothing at $X_i$}
\EndFor
\State Smooth the residuals (Nadaraya--Watson): $\displaystyle \hat r_s(X_i)=\tilde r(X_i)=\frac{\sum_j w_{ij}\,\hat r_j}{\sum_j w_{ij}}$. \Comment{$\mathrm{E}(\tilde r(X_i))=0$, $\mathrm{Var}(\tilde r(X_i))=\dfrac{\sum_j w_{ij}^2}{(\sum_j w_{ij})^2}$}
\State Compute the weight factor (inverse variance of the smoothed residual): $\displaystyle v(X_i)=\frac{\big(\sum_j w_{ij}\big)^2}{\sum_j w_{ij}^2}$. \Comment{makes each observation contribute equally under $H_0$; essential for stable asymptotics}
\State \textbf{Test statistic:} $\displaystyle \hat T=\frac1n\sum_{i=1}^n \hat r_s(X_i)^2\,v(X_i)$. \Comment{scalar form, Eq.~(3.2)}
\State \textbf{Estimation correction --- mean.} Build the hat-type matrix $H=\hat V X^{*}\big[(X^{*})^{\top}\hat V X^{*}\big]^{-1}(X^{*})^{\top}$, using $Y-\hat g(X)\approx (I-H)(Y-g_0(X))$. The corrected null mean is
\Statex $\displaystyle \mathrm{E}(\hat T)=\frac1n\sum_i W_i^{\top}\big(I-\hat V^{1/2}X^{*}[(X^{*})^{\top}\hat V X^{*}]^{-1}(X^{*})^{\top}\hat V^{1/2}\big)W_i,$
\Statex where $W_i$ is the $i$-th column-vector with $j$-th element $w_{ij}/(\sum_k w_{ik}^2)^{1/2}$. \Comment{Eq.~(6.5); shrinks $\mathrm{E}(T)=1$ downward}
\State \textbf{Variance.} Compute the exact estimation-corrected variance
\Statex $\displaystyle \widehat{\mathrm{Var}}(\hat T)=\frac1{n^{2}}\sum_i\sum_j\Big(\textstyle\sum_k w_{ik}^2\sum_k w_{jk}^2\Big)^{-1}\!\Big[\sum_k C_{ik}C_{jk}\,\frac{6\hat\pi_k^2-6\hat\pi_k+1}{\hat\pi_k(1-\hat\pi_k)}+2\big(\textstyle\sum_k C_{ik}C_{jk}\,\hat\pi_k(1-\hat\pi_k)\big)^2\Big],$
\Statex with $C_{ik}=\dfrac{w_{ik}}{\sqrt{\hat\pi_k(1-\hat\pi_k)}}-\sum_s \dfrac{w_{is}\,h_{ns_k}}{\sqrt{\hat\pi_s(1-\hat\pi_s)}}$ (the estimation-adjusted coefficients). \Comment{Eq.~(6.6); without estimation, $C_{ik}\!\to\!w_{ik}$ and the $2(\cdot)^2$ term is $2(\sum_k w_{ik}w_{jk})^2$, giving Eq.~(3.1)}
\State \textbf{Reference distribution.} Match moments of $\hat T$ to a scaled chi-square $c\,\chi^2_\nu$: set
\Statex $\displaystyle \nu=\frac{2\,\mathrm{E}(\hat T)^2}{\widehat{\mathrm{Var}}(\hat T)},\qquad c=\frac{\widehat{\mathrm{Var}}(\hat T)}{2\,\mathrm{E}(\hat T)},\qquad b=\frac1c=\frac{2\,\mathrm{E}(\hat T)}{\widehat{\mathrm{Var}}(\hat T)}.$
\State \textbf{$p$-value.} $\ p=\Pr\!\big[\chi^2_\nu \ge b\,\hat T\big]$ (scaled-chi-square route), or $p=2\big(1-\Phi(|z|)\big)$ with $z=(\hat T-\mathrm{E}(\hat T))/\sqrt{\widehat{\mathrm{Var}}(\hat T)}$ (normal route; scaled-$\chi^2$ preferred at small $\alpha$).
\State \Return reject $H_0$ (lack of fit) if $p<\alpha$, else fail to reject. \Comment{signed terms $\mathrm{sign}(\hat r_s(X_i))\,\hat r_s(X_i)^2 v(X_i)$ serve as per-observation lack-of-fit diagnostics}
\end{algorithmic}
\end{breakablealgorithm}

\needspace{4\baselineskip}
\begin{breakablealgorithm}
\caption{\textbf{Stute--Zhu cumulative-residual (CUSUM) test} \parencite{stute2002model} --- an omnibus, binning-free lack-of-fit check for a fitted logistic generalized linear model (GLM). The statistic is a Cram\'er--von Mises functional $T_{SZ}=\frac{1}{n^2}\sum_{k=1}^{n}C_k^2$ of the residual process cumulated along the estimated linear predictor; its intractable, data-dependent null law is calibrated by a parametric (model-based) bootstrap.}
\label{alg:tsz_bootstrap}
\begin{algorithmic}[1]\small
\Require Data $\{(y_i,X_i)\}_{i=1}^{n}$ with $y_i\in\{0,1\}$ and covariate vector $X_i$; number of bootstrap replications $B$ (e.g.\ $200$--$1000$).
\Ensure Bootstrap $p$-value for $H_0$: the logistic mean model $\pi(x)=m(\beta^{T}x)=\{1+e^{-\beta^{T}x}\}^{-1}$ is correctly specified.

\Statex \textbf{Part 1: Observed statistic $T_{SZ}^{obs}$}
\State Fit the logistic model under $H_0$ by MLE; obtain $\hat{\beta}$.
\State For each $i$ compute the linear predictor (single projection direction) $\hat{\eta}_i \gets \hat{\beta}^{T}X_i$ and the fitted probability $\hat{\pi}_i \gets m(\hat{\eta}_i)=\{1+e^{-\hat{\eta}_i}\}^{-1}$.
\State Compute the raw response residuals $\hat{r}_i \gets y_i-\hat{\pi}_i$. \Comment{$E[\hat r_i]=0$ under $H_0$}
\State Sort observations by the linear predictor: let $(1),(2),\dots,(n)$ be the permutation with $\hat{\eta}_{(1)}\le\hat{\eta}_{(2)}\le\dots\le\hat{\eta}_{(n)}$, giving ordered residuals $\hat{r}_{(i)}$.
\State Form the cumulative residual (CUSUM) process at each ordered point: $C_k \gets \sum_{i=1}^{k}\hat{r}_{(i)}$, for $k=1,\dots,n$.
\Statex \hskip\algorithmicindent \emph{$C_k$ is the discrete version of the marked empirical process}
\Statex \hskip\algorithmicindent $R_n^{*}(u)=n^{-1/2}\sum_{i=1}^{n}\mathbf{1}\{\hat{\beta}^{T}X_i\le u\}\,(y_i-\hat{\pi}_i)$ \emph{of Stute \& Zhu (2002); systematic}
\Statex \hskip\algorithmicindent \emph{misfit makes $C_k$ drift away from $0$.}
\State Compute the Cram\'er--von Mises statistic $\displaystyle T_{SZ}^{obs} \gets \frac{1}{n^{2}}\sum_{k=1}^{n}C_k^{2}$, the discrete analogue of $\int R_n^{*}(u)^2\,dF_n(u)$ with $F_n$ the empirical c.d.f.\ of the $\hat{\eta}_i$.

\Statex \textbf{Part 2: Model-based (parametric) bootstrap null distribution}
\For{$b=1$ to $B$}
    \State Draw a bootstrap response vector $Y^{*}=(y_1^{*},\dots,y_n^{*})$ with $y_i^{*}\sim\mathrm{Bernoulli}(\hat{\pi}_i)$ independently, holding covariates $X_i$ fixed. \Comment{resample under $H_0$}
    \State Refit the same logistic model to $(Y^{*},X)$ to obtain $\hat{\beta}^{*}$, fitted values $\hat{\pi}_i^{*}$, predictors $\hat{\eta}_i^{*}=\hat{\beta}^{*T}X_i$, and residuals $\hat{r}_i^{*}=y_i^{*}-\hat{\pi}_i^{*}$.
    \State Re-sort by $\hat{\eta}_{(i)}^{*}$, cumulate $C_k^{*}=\sum_{i=1}^{k}\hat{r}_{(i)}^{*}$, and compute $\displaystyle T_{SZ}^{*(b)} \gets \frac{1}{n^{2}}\sum_{k=1}^{n}C_k^{*2}$ exactly as in Part 1.
\EndFor

\Statex \textbf{Part 3: $p$-value}
\State $\displaystyle p \gets \frac{1}{B}\sum_{b=1}^{B}\mathbf{1}\{T_{SZ}^{*(b)}\ge T_{SZ}^{obs}\}$; reject $H_0$ at level $\alpha$ if $p\le\alpha$. \Comment{large $T_{SZ}$ signals lack of fit}
\end{algorithmic}
\end{breakablealgorithm}

\needspace{4\baselineskip}
\begin{breakablealgorithm}
\caption{\textbf{GiViTI calibration test and belt} \parencite{nattino2014new, nattino2015new} --- a data-driven \emph{polynomial} calibration goodness-of-fit test for a fitted logistic model. It flexibly regresses the observed logit on a forward-selected polynomial of the model logits and compares the fit to perfect calibration via the likelihood-ratio statistic $T_m=2[\hat{\mathcal{L}}_m-\hat{\mathcal{L}}(\boldsymbol{\alpha}_1)]$, referred to the \emph{non-standard} null CDF $F_m$ induced by the selection of $m$ (\emph{not} a plain $\chi^2_{m+1}$).}
\label{alg:giviti}
\begin{algorithmic}[1]\small
\Require binary outcomes $\boldsymbol{o}=(o_1,\dots,o_n)$, $o_j\in\{0,1\}$; predicted probabilities $\boldsymbol{e}=(e_1,\dots,e_n)$ from the model under assessment; validation mode $\textsc{devel}\in\{\text{external},\text{internal}\}$; forward-selection threshold $q$ (typ.\ $q=0.95$)
\State \textbf{Transform predictions to the logit scale:}\quad $g_j \gets \operatorname{logit}(e_j)=\ln\!\dfrac{e_j}{1-e_j},\qquad j=1,\dots,n$
\Statex \Comment{$g_j$ is the linear predictor of the assessed model; the calibration model is fitted \emph{in} $g_j$.}
\State \textbf{Define the calibration (polynomial logistic) model of degree $m$:}
\Statex \quad $\displaystyle p=f_m(\boldsymbol{\alpha},e)=\operatorname{logit}^{-1}\!\Big(\sum_{i=0}^{m}\alpha_i\,g^{\,i}\Big)=\frac{1}{1+\exp\!\big(-\sum_{i=0}^{m}\alpha_i g^{\,i}\big)}$
\Statex \Comment{$\boldsymbol{\alpha}=(\alpha_0,\dots,\alpha_m)$ fitted by ordinary logistic regression of $o_j$ on $1,g_j,g_j^2,\dots,g_j^m$. Perfect calibration $p=e$ corresponds to $\boldsymbol{\alpha}_1=(0,1,0,\dots,0)$.}
\State \textbf{Set the starting degree of forward selection:}
\If{$\textsc{devel}=\text{external}$} \Comment{frozen model validated on an \emph{independent} sample}
    \State $m_{\text{start}}\gets 1$ \Comment{linear recalibration $\alpha_0+\alpha_1 g$ can be non-trivial; $T_1\sim\chi^2_2$}
\Else \Comment{internal / goodness-of-fit: same data used to fit the model}
    \State $m_{\text{start}}\gets 2$
    \Statex \Comment{MLE optimality forces the linear fit to be exact: $\hat\alpha_0=0,\hat\alpha_1=1$, so $\hat{\mathcal{L}}_1=\hat{\mathcal{L}}(\boldsymbol{\alpha}_1)$ and $T_1\equiv0$; misfit can only appear from non-linear terms $m\ge2$.}
\EndIf
\State \textbf{Forward selection of the polynomial degree $m$ (likelihood-ratio steps):}
\State $m\gets m_{\text{start}}$; fit $f_{m}$ and record $\hat{\mathcal{L}}_{m}=\hat l_{m}$, the maximized log-likelihood $\displaystyle \hat l_{m}=\sum_{j=1}^{n}\big[o_j\ln f_{m}(\hat{\boldsymbol{\alpha}},e_j)+(1-o_j)\ln(1-f_{m}(\hat{\boldsymbol{\alpha}},e_j))\big]$
\Loop \Comment{try to add the next monomial $\alpha_{m+1}g^{\,m+1}$}
    \State fit $f_{m+1}$; compute the LR increment $\;D_{m}\gets 2\big(\hat l_{m+1}-\hat l_{m}\big)\;\sim\;\chi^2_1$ under $H_0$
    \If{$D_{m}>\chi^2_{1,q}$} \Comment{$\chi^2_{1,q}=F^{-1}_{\chi^2_1}(q)$; term is significant at level $q$}
        \State $m\gets m+1$ \Comment{accept the higher-order term and continue}
    \Else
        \State \textbf{stop the loop} \Comment{first non-significant term; final degree is $m$}
    \EndIf
\EndLoop
\State \textbf{Compute the calibration test statistic (LR vs.\ perfect calibration):}
\Statex \quad $\displaystyle \hat{\mathcal{L}}(\boldsymbol{\alpha}_1)=\sum_{j=1}^{n}\big[o_j\ln e_j+(1-o_j)\ln(1-e_j)\big]$ \Comment{log-likelihood of the ideal model $p=e$}
\Statex \quad $\displaystyle T_m \;=\; 2\big[\hat{\mathcal{L}}_m-\hat{\mathcal{L}}(\boldsymbol{\alpha}_1)\big]$
\Statex \Comment{Equivalently $T_m=T_1+\sum_{i=1}^{m-1}D_i$ (external, $T_1\sim\chi^2_2$) and $T_m=\sum_{i=1}^{m-1}D_i$ (internal, since $T_1=0$).}
\State \textbf{Evaluate the correct null CDF $F_m$ of $T_m$} (it accounts for the data-driven selection of $m$ and is \emph{not} a plain $\chi^2_{m+1}$): closed forms and one-dimensional integral representations of $F_m$ are given by \textcite{nattino2014new} (Eqs.~17--19) for external validation, and by \textcite{nattino2015new} (Eq.~10) for the internal case, where notably $F_2(t)=F_{\chi^2_1}(t)$ (one degree of freedom).
\State \textbf{Report the $p$-value:}\quad $p\text{-value}\gets 1-F_m(T_m)$
\State \textbf{(Belt)} \textbf{Construct the calibration belt at confidence $1-\gamma$:} find $k_\gamma$ with $P\big(2(\hat{\mathcal{L}}_m-\hat{\mathcal{L}}(\boldsymbol{\alpha}_1))<k_\gamma\big)=1-\gamma$ by inverting $F_m$; form $R_\gamma=\{\boldsymbol{\alpha}: 2(\hat{\mathcal{L}}_m-l(\boldsymbol{\alpha}))<k_\gamma\}$ and, for each $e$, plot the band
\Statex \quad $\big(p^{\min}(e),p^{\max}(e)\big)=\Big(\min_{\boldsymbol{\alpha}\in R_\gamma}\operatorname{logit}^{-1}(\boldsymbol{\alpha}^{\!\top}\boldsymbol{g}_e),\;\max_{\boldsymbol{\alpha}\in R_\gamma}\operatorname{logit}^{-1}(\boldsymbol{\alpha}^{\!\top}\boldsymbol{g}_e)\Big),\quad \boldsymbol{g}_e=(1,g_e,\dots,g_e^m)^{\!\top}.$
\Statex \Comment{If the band contains the bisector $p=e$ over the whole range, no significant miscalibration; regions where the belt excludes the bisector localize the direction (over-/under-prediction) of misfit.}
\State \Return $T_m$, $m$, $p\text{-value}$, and the calibration belt.
\end{algorithmic}
\end{breakablealgorithm}

One can notice that the six algorithms share one recipe: from the fitted probabilities $\hat\pi_i$ they build a single statistic and compare it to a known reference. \textbf{Hosmer--Lemeshow}, \textbf{Osius--Rojek} and \textbf{McCullagh} reduce the data to a (standardized) chi-square; \textbf{le~Cessie} and \textbf{Stute--Zhu} summarize the residuals---smoothed or cumulated---and \textbf{GiViTI} compares nested polynomial fits. A few symbols recur across the boxes: $v_i=\hat\pi_i(1-\hat\pi_i)$ is the Bernoulli variance of observation $i$, $\Phi$ is the standard-normal cumulative distribution function, $G$ is the number of groups, and $\mathbf{X}$ is the $n\times p$ design matrix of predictors (distinct from the statistic $\Xsq$). The leverage-type quantity $h_i=\mathbf{x}_i^\top(\mathbf{X}^\top\mathbf{W}\mathbf{X})^{-1}\mathbf{x}_i$ (so the hat-matrix diagonal equals $v_i h_i$) measures how strongly observation $i$ pulls its own fitted value---it is what the standardized-Pearson tests use to correct the variance for having estimated $\boldsymbol\beta$. In \textbf{McCullagh}'s box, $\hat\kappa_{i3}$ and $\hat\kappa_{i4}$ are the third and fourth (skewness- and kurtosis-type) cumulants of the Bernoulli contribution, while $\hat u_i$ are the fitted values and $\mathrm{RSS}_u$ the residual sum of squares of an auxiliary regression of $\mathbf u$ on $\mathbf{X}$; in \textbf{GiViTI}'s, $\hat{\mathcal{L}}_m$ is the maximized log-likelihood of the degree-$m$ polynomial calibration model. (For grouped data, Osius--Rojek's correction $A$ also uses $\bar m$, the harmonic mean of the group sizes; it vanishes in the fully sparse case studied here.)

\section{Reproducing the Simulation Benchmark}
\label{app:repro}
As an example, a single replication of the benchmark (here, the pronounced omitted-interaction scenario, with a continuous and a dichotomous covariate) is run as follows. Looping the same code over the sample sizes, covariate designs, and misspecification settings of Section~\ref{sec:design} reproduces the full Type~I error and power study.
\begin{verbatim}
# install ebrahim.gof -- first time only
install.packages("ebrahim.gof")

# load it
library(ebrahim.gof)

# install dependencies -- first time only
install.packages(c("givitiR", "callr", "BAGofT", "mgcv", "ResourceSelection",
                   "CompQuadForm", "statmod"), dependencies = TRUE)

# one iteration for the pronounced omitted-interaction scenario
set.seed(2024)
n <- 500
x <- runif(n, -3, 3)
d <- rbinom(n, 1, 0.5)                                # dichotomous covariate
eta <- -1.792 + 0.135*x + 1.791*d + 0.597*x*d         # true (pronounced interaction)
y   <- rbinom(n, 1, plogis(eta))
fit <- glm(y ~ x + d, family = binomial)              # misspecified: x:d omitted

## run the full battery of tests reported in the paper
## (may take 10 to 20 minutes):
run.all.gof(fit, tests = "all", include_slow = TRUE,
            control = list(BAGofT = list(nsim = 100)))
\end{verbatim}
The covariate-space tests (Pulkstenis--Robinson, PR-GAM) operate on the covariates rather than on the fitted probabilities, so they require at least one categorical covariate; the adaptive BAGofT test further needs more than one predictor to build its random-forest partition, together with a chosen number of resampling iterations (\texttt{nsim}, which sets the resolution of its empirical $p$-value). These conditions are satisfied here, but not in a single continuous-covariate design, where these tests are reported as not applicable.
\end{appendices}

\backmatter

\bmhead{Acknowledgements}
During the preparation of this manuscript, the authors used Claude (Anthropic) solely to improve the language, clarity, and readability of the text. All AI-assisted output was reviewed and edited by the authors, who take full responsibility for the content. No AI tool was used to generate scientific content, results, or interpretations, and no AI tool is listed as an author.

\section*{Statements and Declarations}

\noindent\textbf{Funding.} No funding was received for conducting this study. Open-access publication is supported by the STDF/EKB--Springer Nature transformative agreement.

\smallskip\noindent\textbf{Competing interests.} The authors have no competing interests to declare that are relevant to the content of this article.

\smallskip\noindent\textbf{Ethics approval.} Not applicable. This study uses only previously published, publicly available benchmark data and computer simulations; no new data involving human participants or animals were collected.

\smallskip\noindent\textbf{Data availability.} The low-birth-weight data (188 records) are the classic Hosmer--Lemeshow data \parencite{hosmer2013applied}, closely related to \code{birthwt} ($n=189$) in the R package \texttt{MASS}. The exact analyzed version, all simulation drivers, aggregated result files, and figure code are permanently archived at \url{https://doi.org/10.5281/zenodo.21286172}; every number and figure in the paper is reproducible from these files.

\smallskip\noindent\textbf{Code availability.} All goodness-of-fit and calibration tests are implemented in the open-source R package \texttt{ebrahim.gof} (available on CRAN); a single call to \code{run.all.gof()} runs the full battery. Development source: \url{https://github.com/ebrahimkhaled/ebrahim.gof}.

\smallskip\noindent\textbf{Author contributions.} E.K.E.\ conceived the study, implemented the software, and designed, ran, and wrote up the experiments and the manuscript. A.E.\ supervised the research. Both authors read and approved the final manuscript.

\bibliographystyle{plainnat}
\bibliography{references}

\end{document}